\documentclass[]{aa}  

\usepackage{graphicx}
\usepackage{txfonts}
\usepackage{adjustbox}
\usepackage{caption}
\usepackage{float}
\usepackage[flushleft]{threeparttable}
\usepackage{pdfpages}
\usepackage{longtable}
\usepackage{textcomp}
\usepackage{multirow}
\usepackage{enumerate}
\usepackage{hhline}

\begin{document}

   \title{The X-ray properties of z$\sim$6 luminous quasars}


   \author{R. Nanni
          \inst{1,2}
          \and C. Vignali
          \inst{1,2}
          \and R. Gilli
          \inst{1}
          \and A. Moretti
          \inst{3}
          \and W. N. Brandt
          \inst{4,5,6}
           }

   \institute{INAF, Osservatorio Astronomico di Bologna, via Gobetti 93/3, 40129 Bologna, Italy
         \and
             Dipartimento di Astronomia, Universit\`a degli Studi di Bologna, via Gobetti 93/2, 40129 Bologna, Italy
          \and
             INAF, Osservatorio Astronomico di Brera, via Brera 28, 20121 Milano, Italy
          \and 
              Department of Astronomy and Astrophysics, 525 Davey Lab, The Pennsylvania State University, University Park, PA 16802, USA
           \and
              Institute for Gravitation and the Cosmos, The Pennsylvania State University, University Park, PA 16802, USA
            \and 
               Department of Physics, 104 Davey Laboratory, The Pennsylvania State University, University Park, PA 16802, USA\\
             }

   \date{}
 
  \abstract
  {We present a systematic analysis of X-ray archival data of all the 29 quasars (QSOs) at $z$ > 5.5 observed so far with \textit{Chandra}, XMM\textit{-Newton} and \textit{Swift}-XRT, including the most-distant quasar ever discovered, ULAS J1120+0641 ($z$ = 7.08). This study allows us to place constraints on the mean spectral properties of the primordial population of luminous Type 1 (unobscured) quasars.
 Eighteen quasars are detected in the X-ray band, and we provide spectral-fitting results for their X-ray properties, while for the others we provide upper limits to their soft (0.5-2.0 keV) X-ray flux.
We measured the power-law photon index and derived an upper limit to the column density for the five quasars (J1306+0356, J0100+2802, J1030+0524, J1148+5251, J1120+0641) with the best spectra (> 30 net counts in the 0.5-7.0 keV energy range) and find that they are consistent with values from the literature and lower-redshift quasars. By stacking the spectra of ten quasars detected by \textit{Chandra} in the redshift range 5.7 $\le$ $z$ $\le$ 6.1 we find a mean X-ray power-law photon index of $\Gamma = 1.92_{-0.27}^{+0.28}$ and a neutral intrinsic absorption column density of $N_H \le 10^{23}$ cm$^{-2}$. These results suggest that the X-ray spectral properties of luminous quasars have not evolved up to $z$ $\approx$ 6. We also derived the optical-X-ray spectral slopes ($\alpha_{ox}$) of our sample and combined them with those of previous works, confirming that $\alpha_{ox}$ strongly correlates with UV monochromatic luminosity at 2500 \AA . These results strengthen the non-evolutionary scenario for the spectral properties of luminous active galactic nuclei (AGN). 
  }
  
   \keywords{quasars - active galactic nuclei - X-ray - high redshift
               }

   \maketitle
%

\section{Introduction}

Active galactic nuclei (AGN) are one of the best probes of the primordial Universe at the end of the dark ages.
Studying the properties of $z$ $\sim$ 6 quasars is important to understand the formation and early evolution of supermassive black holes (SMBHs) and their interaction with the host galaxy.
The presence of SMBHs, $10^8$ - $10^9$ $M_{\odot}$, observed in quasars (QSOs) up to $z$ = 6-7 (e.g., \citealt{Fan06}; \citealt{Ban16}), and hence formed in less than 1 Gyr, is a challenge for modern astrophysics. In order to explain these SMBH masses, accretion of gas must have proceeded almost continuously close to the Eddington limit with fairly low radiative efficiency ($\eta$ < 0.1). The seeds of the observed SMBHs could either be the remnants of PopIII stars (100 $M_{\odot}$; e.g., \citealt{MR01}), or more massive ($10^{4-6}$ $M_{\odot}$) BHs formed from the direct collapse of primordial gas clouds (e.g., \citealt{Vol08}). In the case of lower-mass seeds (PopIII stars), super-Eddington accretion is likely required to form the black-hole masses of $z$ $\sim$ 6 QSOs (e.g., \citealt{Mad14}; \citealt{Vol16}; \citealt{Pez17}).
   
As of today, 198 QSOs have been discovered at redshift $z$ > 5.5 with wide-area optical and IR surveys (e.g.,  \citealt{Fan06}; \citealt{Wil10}; \citealt{Ven13}; \citealt{Mat16}; \citealt{Ban16}; \citealt{Tan16}; \citealt{Yan17}). In particular, wide-area near-IR surveys are now pushing the QSO redshift frontier to $z$ > 6.4. Eight of the 198 QSOs were selected using Spectral Energy Distribution (SED) model fitting to photometric data, and then spectroscopically confirmed (\citealt{Ree17}).
Only a few of these 198 QSOs have been studied through their X-ray emission (e.g., \citealt{Bra02}; \citealt{Far04}; \citealt{Vig05}; \citealt{She06}; \citealt{Mor14}; \citealt{Pag14}; \citealt{Ai16}).
These studies showed that the X-ray spectral properties of high-redshift quasars do not differ significantly from those of AGN at lower redshift. This is generally consistent with observations showing that the broad-band SEDs and the rest-frame IR/optical/UV spectra of quasars have not significantly evolved over cosmic time (e.g., \citealt{Mor11}; \citealt{Bar15}), with a few notable exceptions for the IR band (e.g., \citealt{Jia10}).
   
In this work we provide a systematic analysis of all X-ray data available for the 29 out to 198 QSOs, that were observed by \textit{Chandra}, XMM\textit{-Newton}, and \textit{Swift}-XRT in order to derive the general properties of accretion onto SMBHs at very high redshift. While the X-ray spectral properties of $z$ < 5 quasars are now well established, the situation for quasars at the highest redshifts is not so clear.
In our work we present the most up-to-date and complete X-ray study of the population of quasars in the redshift range 5.5 $\le$ $z$ $\le$ 7.1 by which we managed to place constraints on the X-ray properties of primordial AGN.

The paper is organized as follows. In \S 2 we describe the\linebreak X-ray archival data and their reduction procedure. The data analysis is presented in \S 3, where we also provide a detailed spectral study for those sources with higher photon statistics (> 30 net counts, i.e.,  background-subtracted, in the 0.5-7.0 keV energy band). In \S 4 we discuss the mean X-ray properties of our sample, and in \S 5 we provide estimates of the optical-X-ray spectral slope. In \S 6 we give a summary of our results. Throughout this paper we assume $H_0 = 70$ km s$^{-1}$ Mpc$^{-1}$, $\Omega_{\Lambda} = 0.7$, and $\Omega_M = 0.3$ (\citealt{Ben13}).
\section{Sample selection and data reduction}

To study the X-ray properties of the population of AGN at high-redshift ($z$ > 5.5) we started from the most up-to-date compilation of 198 luminous high-redshift quasars (181 from \citealt{Ban15} \& \citealt{Ban16}; eight from \citealt{Ree17}; nine from \citealt{Yan17}) and cross-correlated it with all the available archival data from \textit{Chandra}, XMM\textit{-Newton}, and \textit{Swift}-XRT. The majority of these 198 AGN were spectroscopically identified with optical and NIR surveys and are classified as Type 1 AGN.
From the cross-correlation we found that 29 sources have archival X-ray observations: 21 QSOs have been observed by \textit{Chandra}, while 12 have XMM\textit{-Newton} observations; J0100+2802, J1030+0524, J1120+0641, J1148+5253 and J1148+5251 were observed by both telescopes. One additional source has been observed by \textit{Swift}-XRT with a relatively deep exposure.
We also note that a further ten objects fall within \textit{Swift}-XRT fields observed for only $\sim$5 ks each. We did not consider them in this work as no useful constraints can be derived on their X-ray properties.
None of these 29 sources come from either the Chandra Deep Field North (\citealt{Xue16}) or South (\citealt{Luo17}) or from the COSMOS survey (\citealt{Civ16}). These three deep fields have no sources with spectroscopic redshift above 5.5 (see \citealt{Vit13} and \citeyear{Vit16} for the Chandra Deep Fields, and \citealt{Mar16} for the COSMOS survey). More generally, there are no X-ray selected sources with spectroscopic redshift > 5.5.

We reprocessed all the 21 \textit{Chandra} sources using the \textit{Chandra} software CIAO v. 4.8 with \textit{faint} or \textit{vfaint} mode for the event telemetry format according to the corresponding observation. Data analysis was carried out using only the events with ASCA grades 0, 2, 3, 4 and 6.
We extracted the number of counts from circular regions centered at the optical position of every source. We used a radius of 2", corresponding to 95\% of the encircled energy fraction (EEF) at 1.5 keV for the on-axis cases ($\theta < 1$'), and of 10" for the off-axis extractions, corresponding to at least 90\% of the EEF at 1.5 keV.
Fifteen of the 21 \textit{Chandra} QSOs were the targets of the X-ray observations, while the other six were serendipitously observed at large off-axis angles ($\theta$ > 1').
The background spectra were extracted from adjacent circular regions, free of sources, with an area ten times larger.
In order to assess if a source could be considered detected in the X-ray band we computed the Poisson probability (P$_P$) of reproducing a number of counts equal to or above the value extracted in the source region (in the 0.5-7.0 keV energy range) given the background counts expected in the source area. We considered as detected those sources showing a detection probability of > 99.7\% (> 3$\sigma$).
 We found that the 15 on-axis QSOs are detected (P$_P$ > 3$\sigma$) in the 0.5-7.0 keV X-ray band. One of the six off-axis sources (RD J1148+5253) is also detected in the X-ray band with low-statistics ($\sim$3 counts; see \S 3.2 of \citealt{Gal17} for detailed investigation of the detection significance) so, in the end, we found that 16 out of 21 sources (including J1148+5253) are detected.
  
The XMM EPIC data were processed using the Science Analysis Software (SAS v. 15) and filtered for high-background time intervals; for each observation and camera we extracted the 10-12 keV light curves and filtered out the time intervals where the light curve was 3$\sigma$ above the mean.
For the scientific analysis we considered only events corresponding to patterns 0-12 and patterns 0-4 for the MOS1/2 and pn, respectively.
Because of the higher background level of XMM, we extracted the counts from circular regions centered at the optical position of the QSOs with radius of 10" for on-axis sources, corresponding to 55\% of EEF at 1.5 keV, and of 30" for off-axis positions, corresponding to at least 40\% of the EEF at 1.5 keV.
The background was extracted using the same approach adopted for \textit{Chandra} data. 
We then computed the Poisson detection probability, as we did for the \textit{Chandra} quasars, for all the sources.
In this case we found that the five on-axis sources (the targets of the corresponding observations) were detected in the X-ray band at > 3$\sigma$, while seven sources were observed with large off-axis angles and are undetected in the X-ray band (they are serendipitously observed).

For the source observed by \textit{Swift}-XRT, data reduction and spectrum extraction were performed using the standard software (HEADAS software v. 6.18)
 and following the procedures described in the instrument user guide.\footnote{http://heasarc.nasa.gov/docs/swift/analysis/documentation}
Given the limited number of photons, in order to optimize the ratio between signal and background we restricted our analysis
to a circular region of 10" radius, including $\sim$50\% of the flux according to the instrumental point spread function (PSF) full width half maximum (FWHM) (\citealt{Mor05}).
The ancillary response file (ARF) has been calculated accordingly by the {\it  xrtmkarf} task.
 In Table 1 we report all the information linked to the X-ray observations of the 29 QSOs. We show in Figure 1 the redshift distribution of all the 198 QSOs known at $z$ > 5.5 (black histogram) and the
distribution of those observed in the X-rays (red shaded histogram). The blue shaded histogram shows the redshift distribution of the 18 QSOs detected. 
We display the X-ray cutouts of the 18 detected sources in Figure 2.
\begin{figure}
 \includegraphics[height=8cm, width=8cm, keepaspectratio]{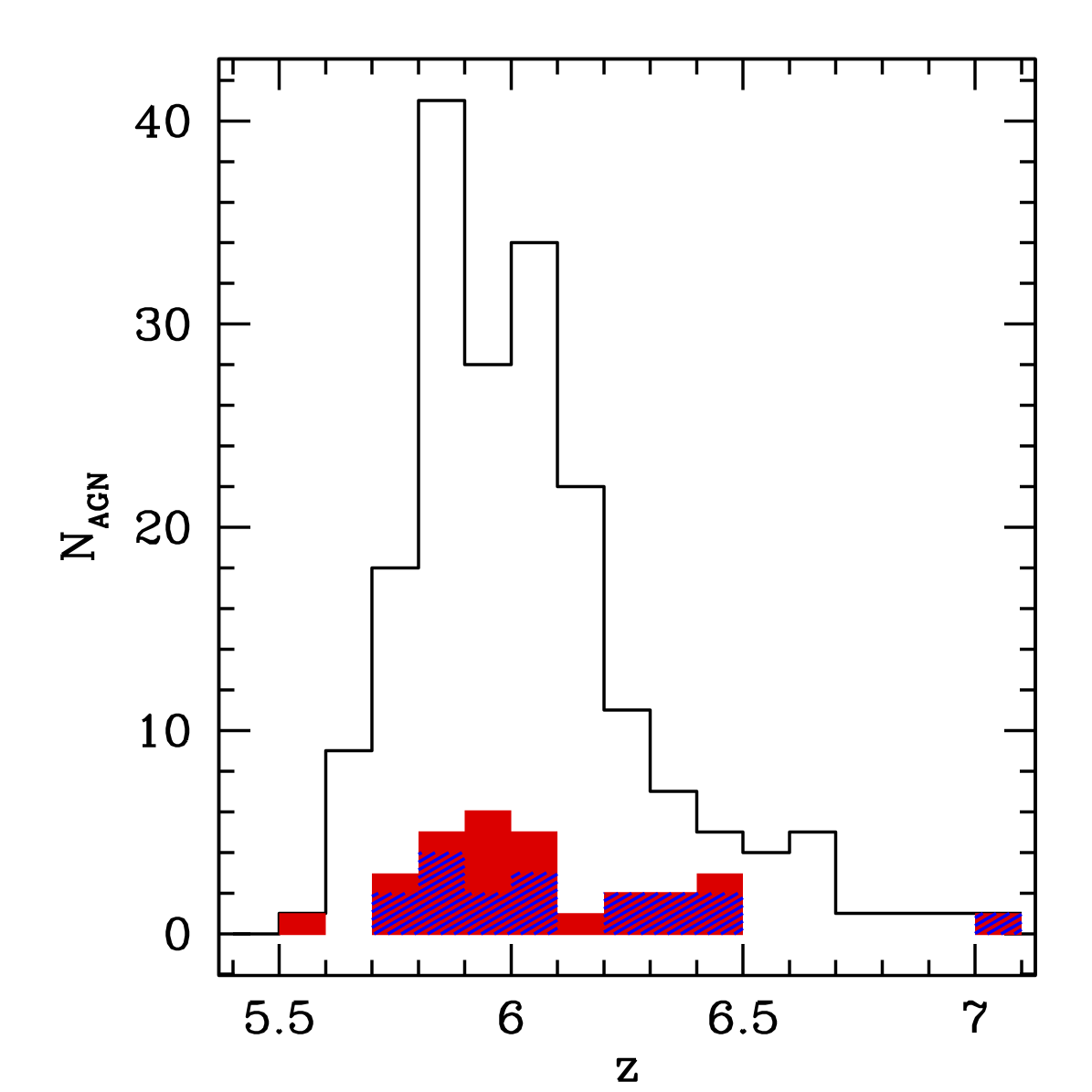}
 \caption{Redshift distribution of the 198 known QSOs at $z$ > 5.5 (black histogram) and of the 29 QSOs observed in the X-rays (red shaded histogram). The blue shaded histogram shows the distribution of the 18 sources detected at > 3$\sigma$. }
\end{figure}
\begin{figure*}
\centering
 \includegraphics[height=4cm, width=4cm, keepaspectratio]{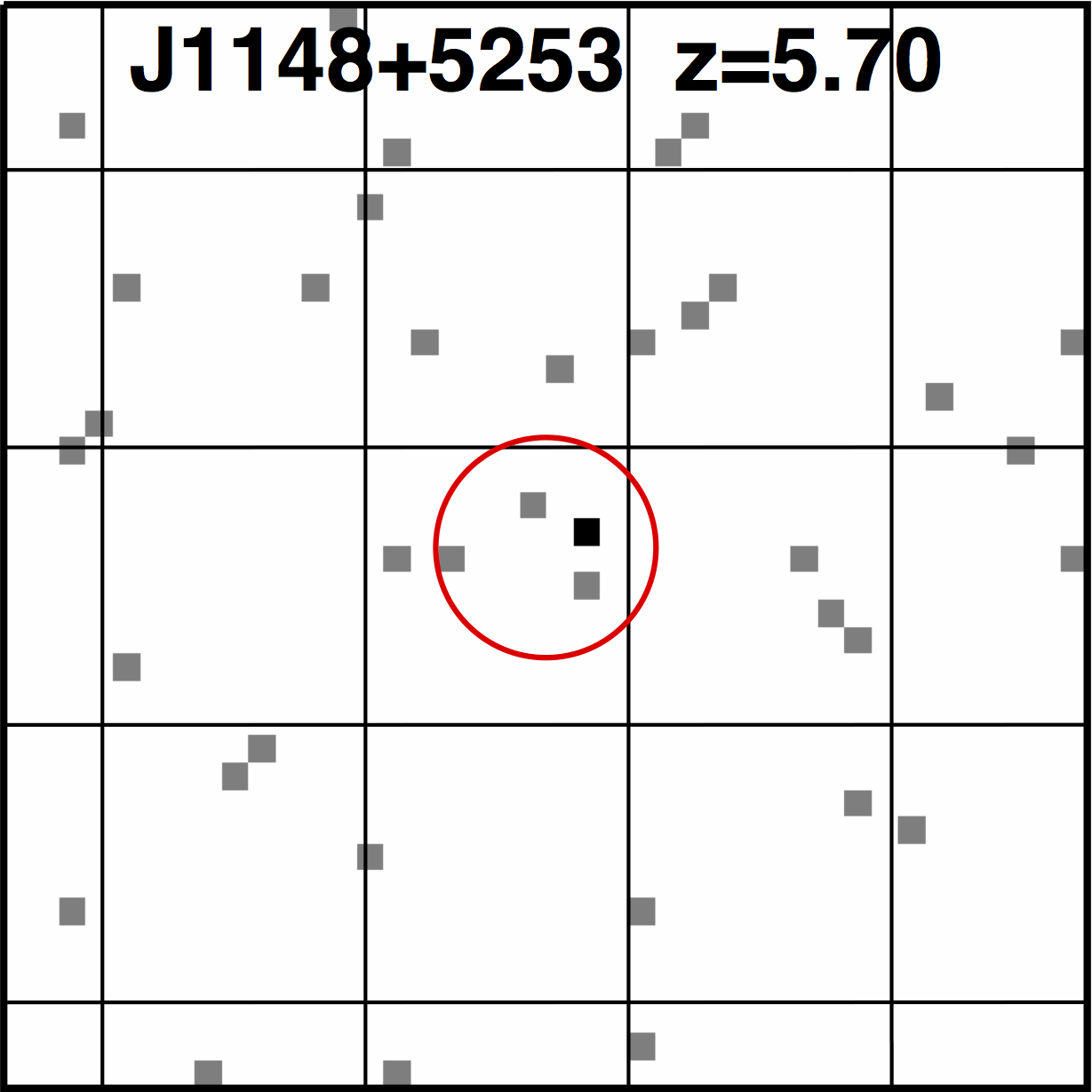} \quad \includegraphics[height=4cm, width=4cm, keepaspectratio]{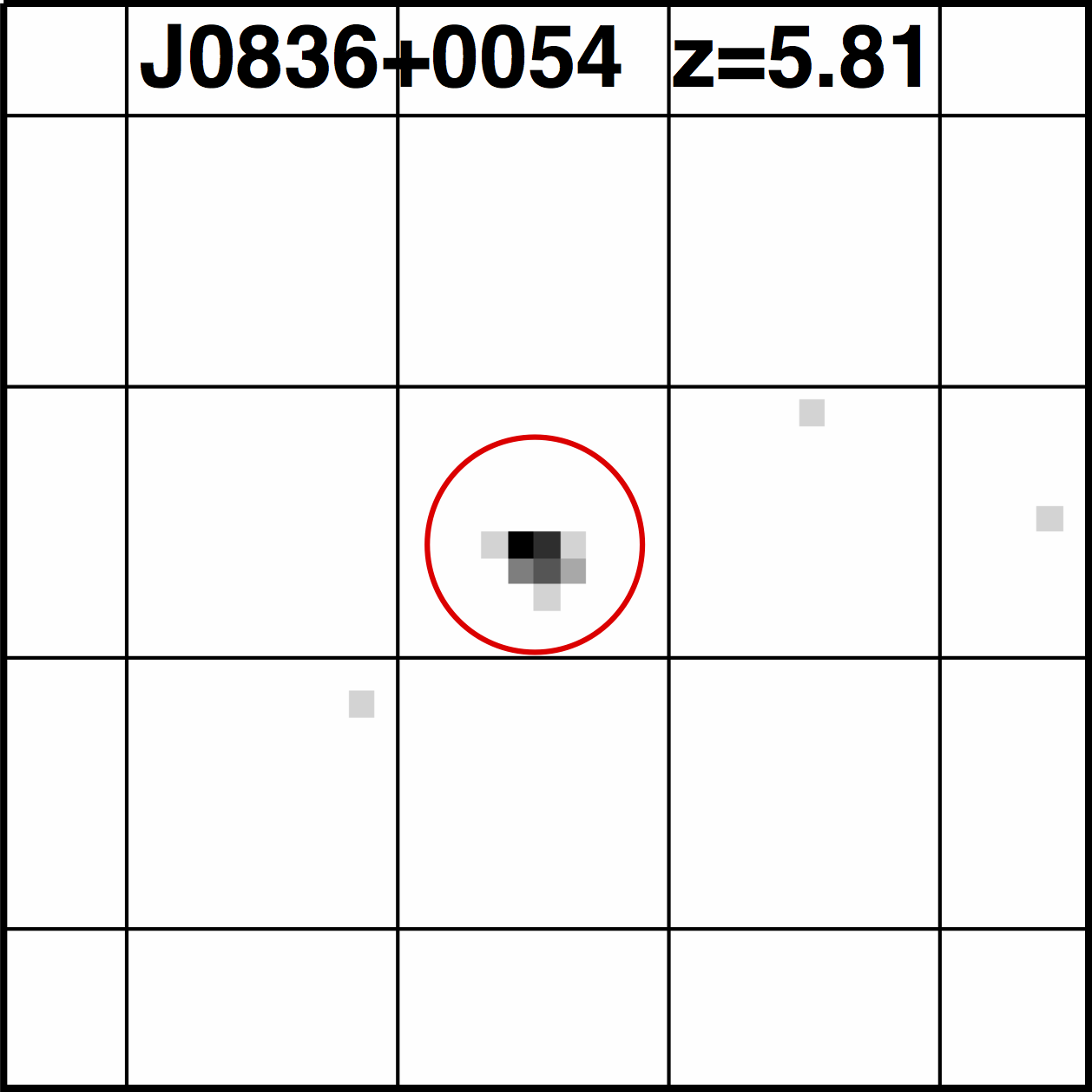}
 \vspace{0.1cm}
 \quad \includegraphics[height=4cm, width=4cm, keepaspectratio]{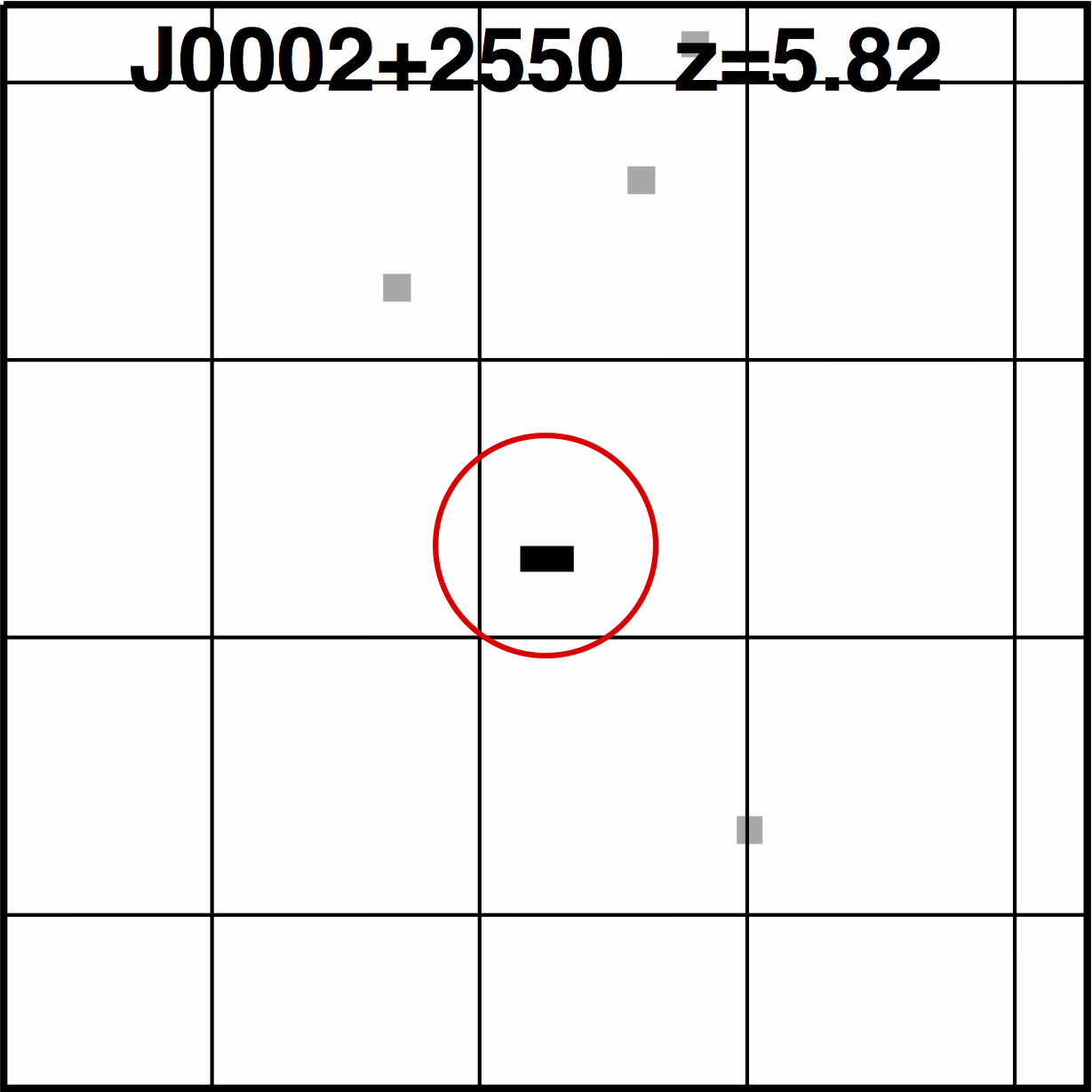} \quad \includegraphics[height=4cm, width=4cm, keepaspectratio]{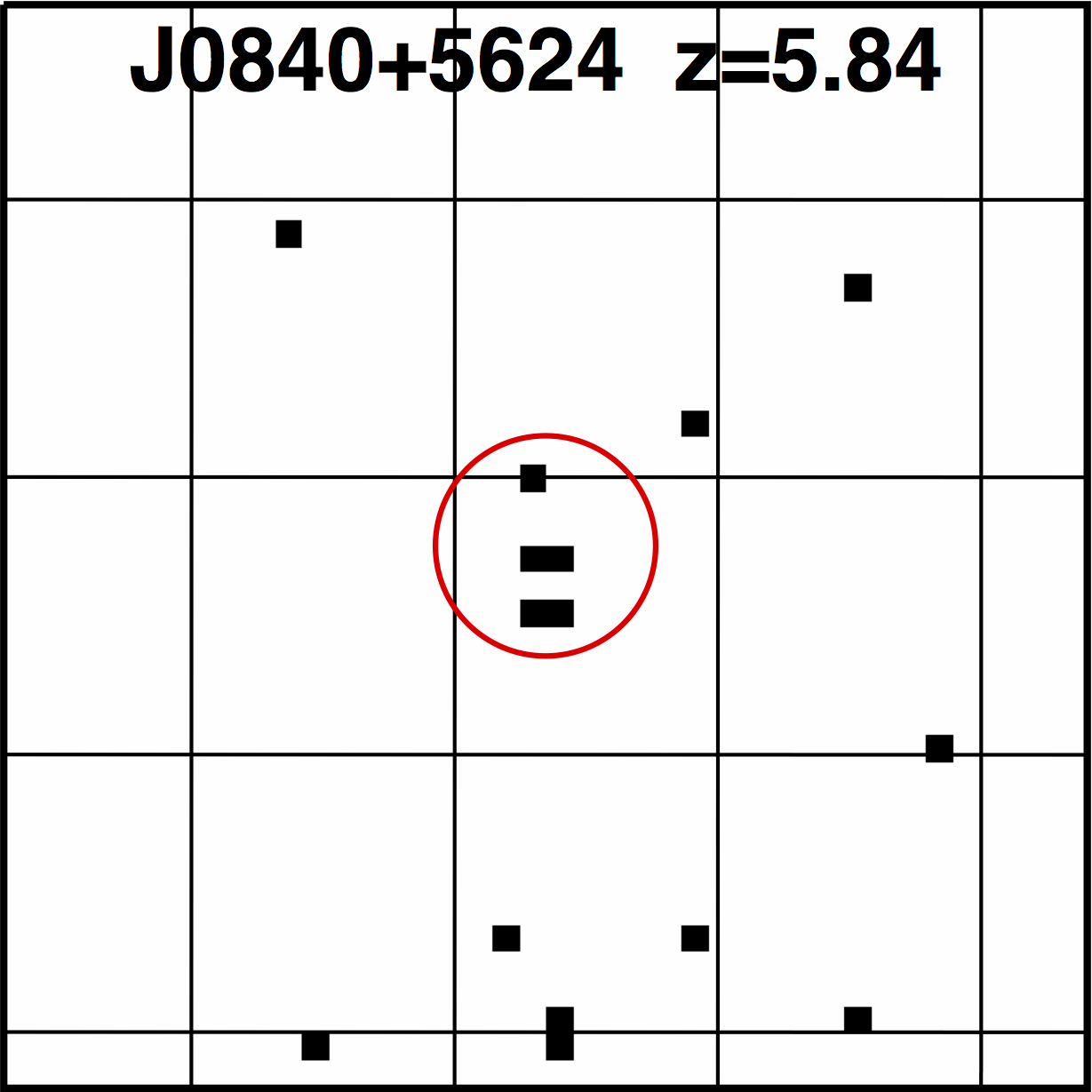}
 \vspace{0.1cm}
 \includegraphics[height=4cm, width=4cm, keepaspectratio]{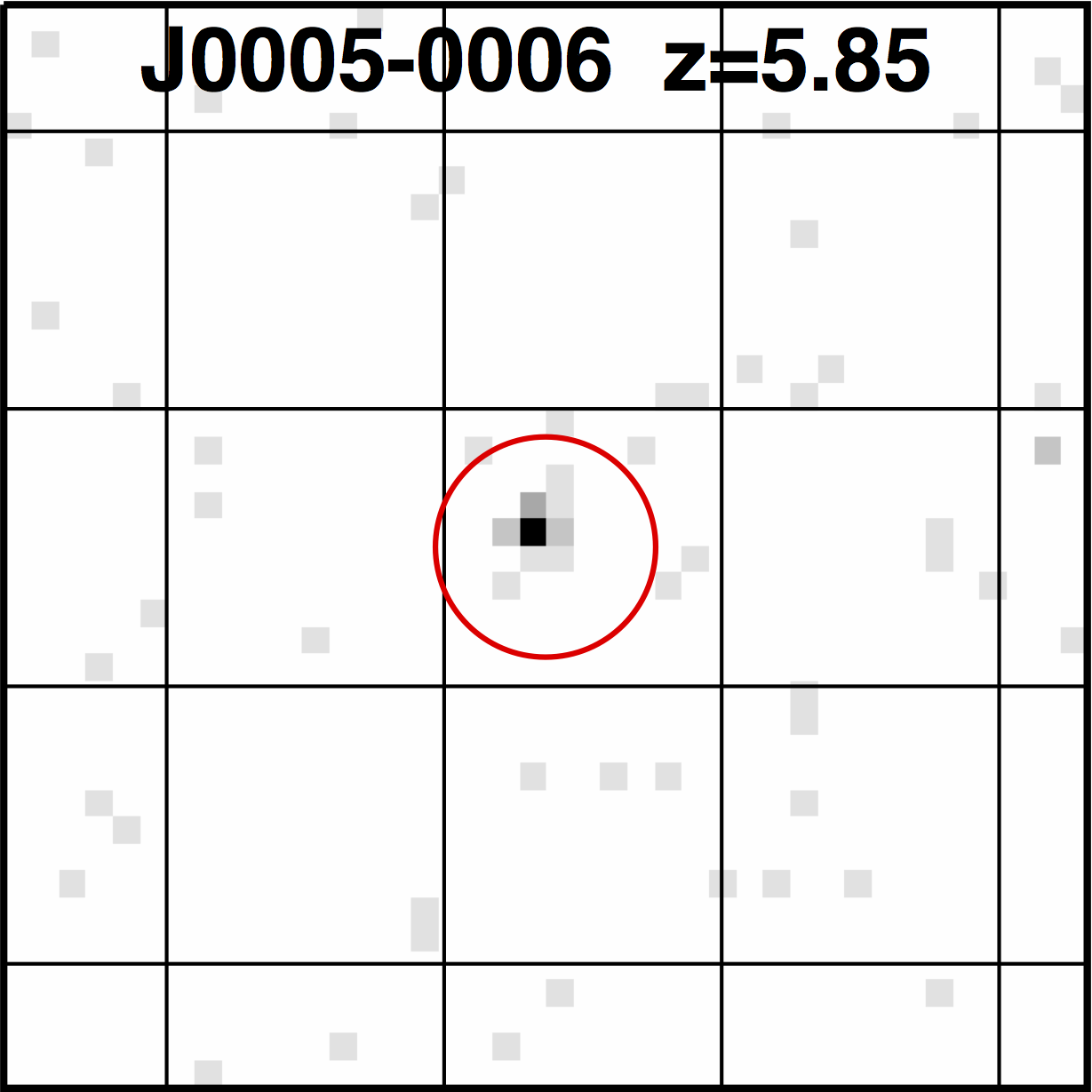} \quad \includegraphics[height=4cm, width=4cm, keepaspectratio]{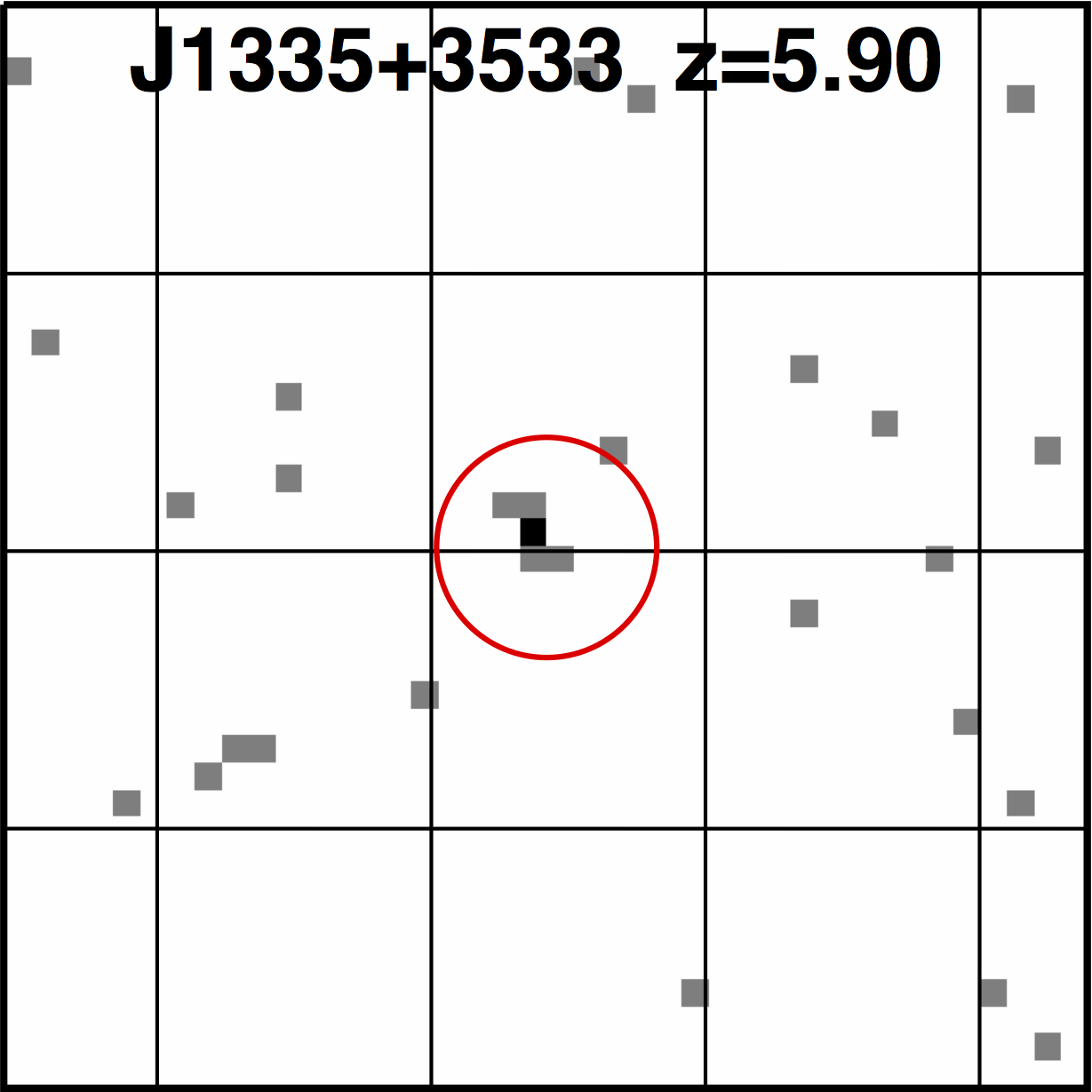}
 \quad \includegraphics[height=4cm, width=4cm, keepaspectratio]{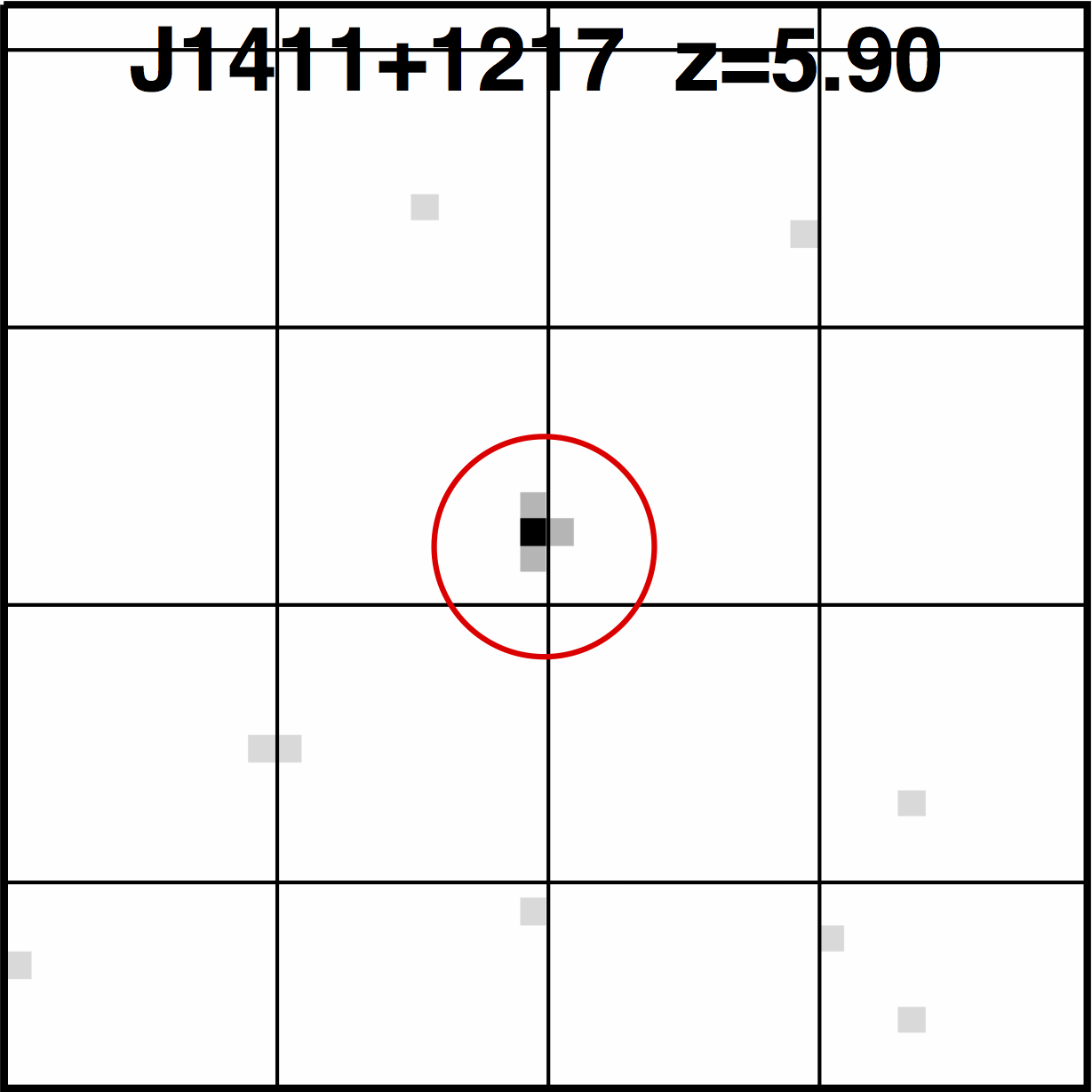} \quad \includegraphics[height=4cm, width=4cm, keepaspectratio]{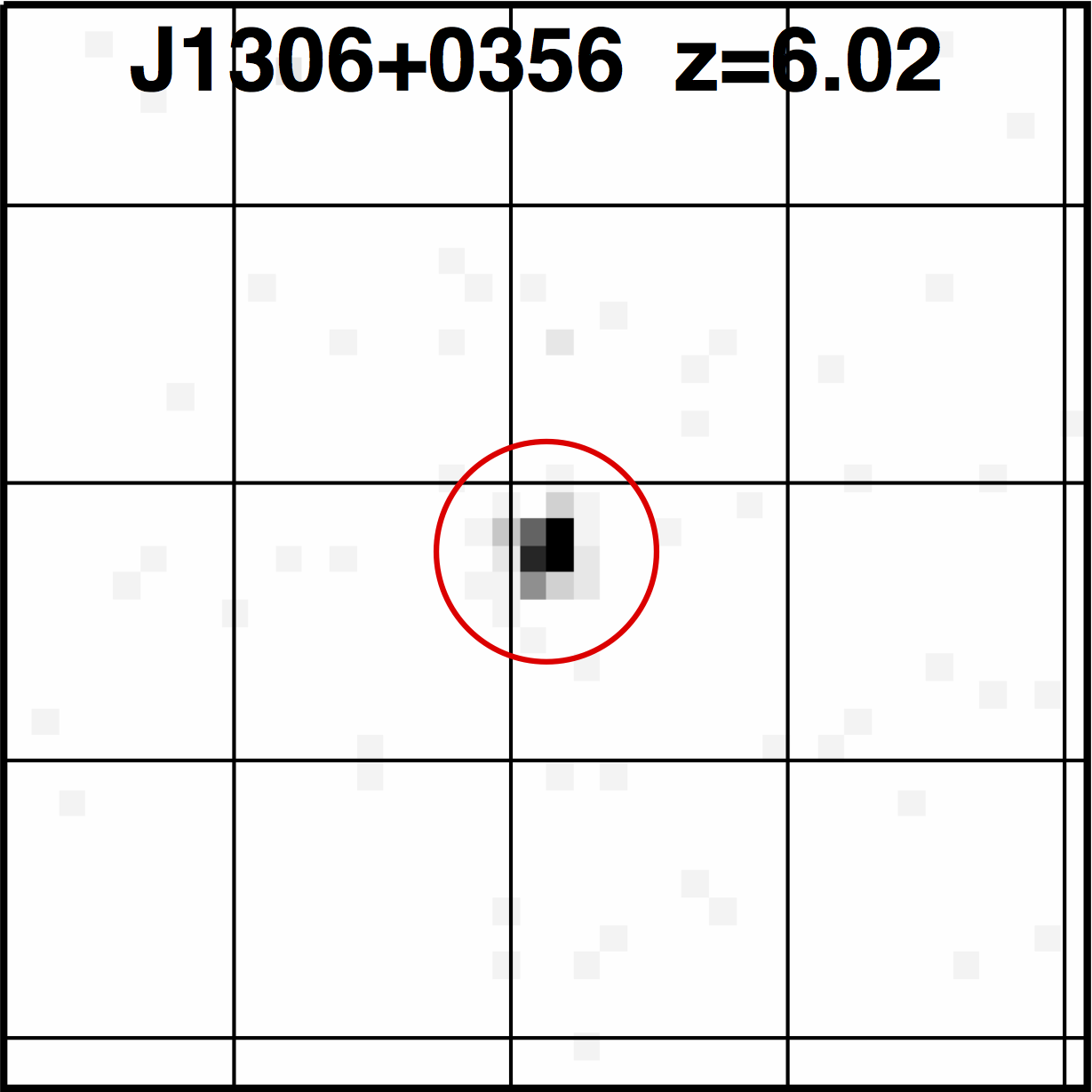}
 \vspace{0.1cm}
 \includegraphics[height=4.005cm, width=4cm]{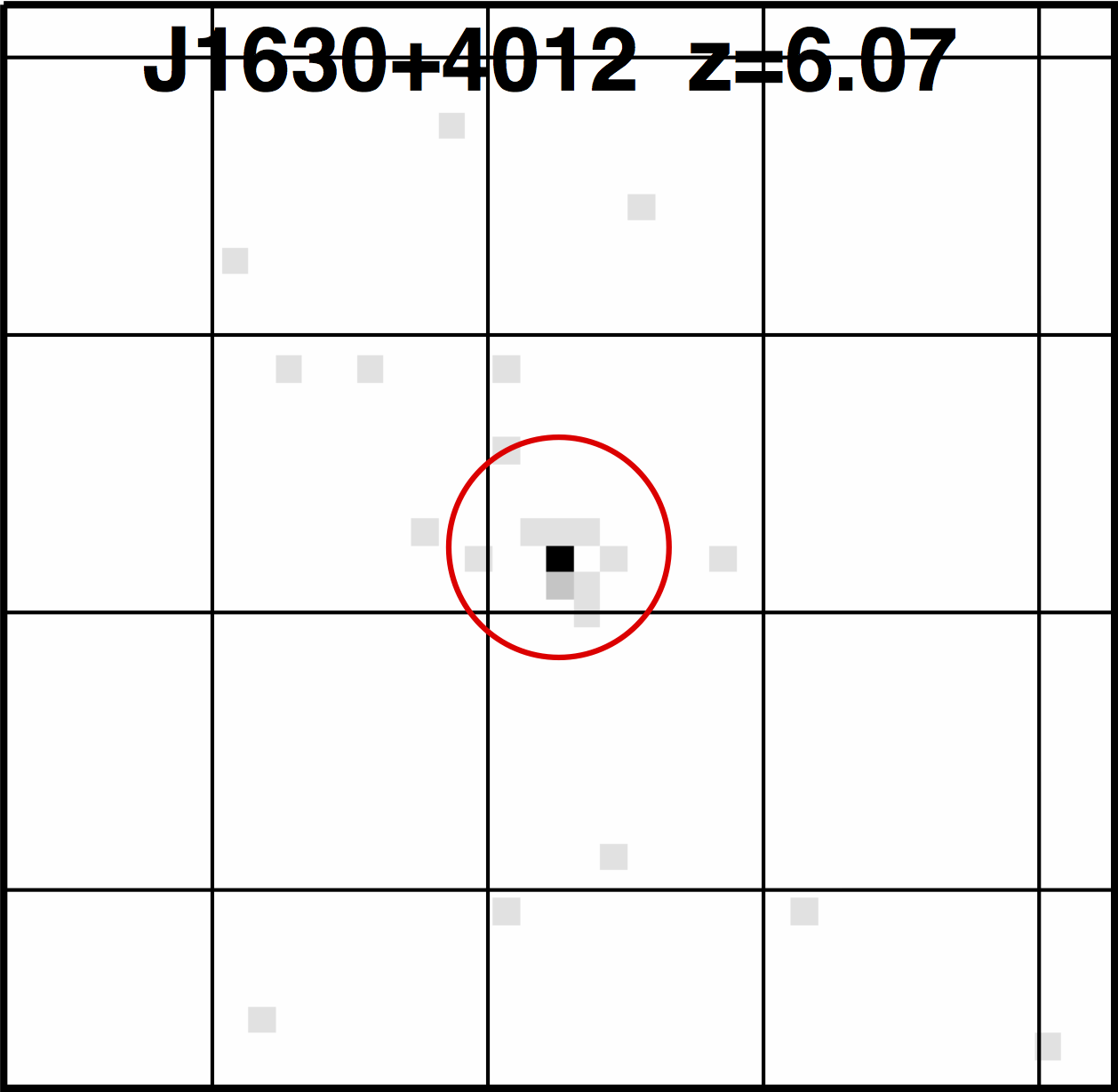} \quad \includegraphics[height=4cm, width=4cm, keepaspectratio]{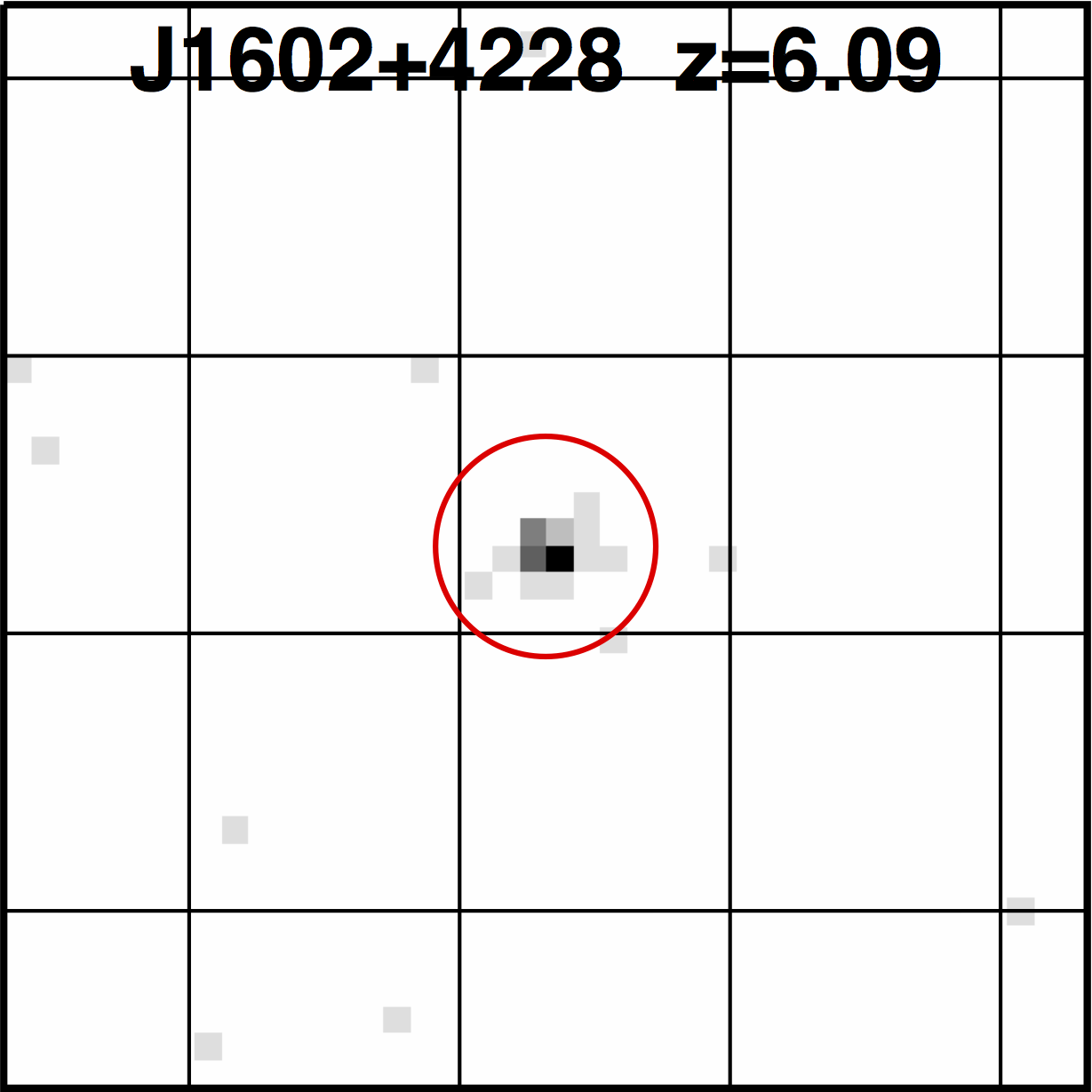}
 \quad \includegraphics[height=4cm, width=4cm, keepaspectratio]{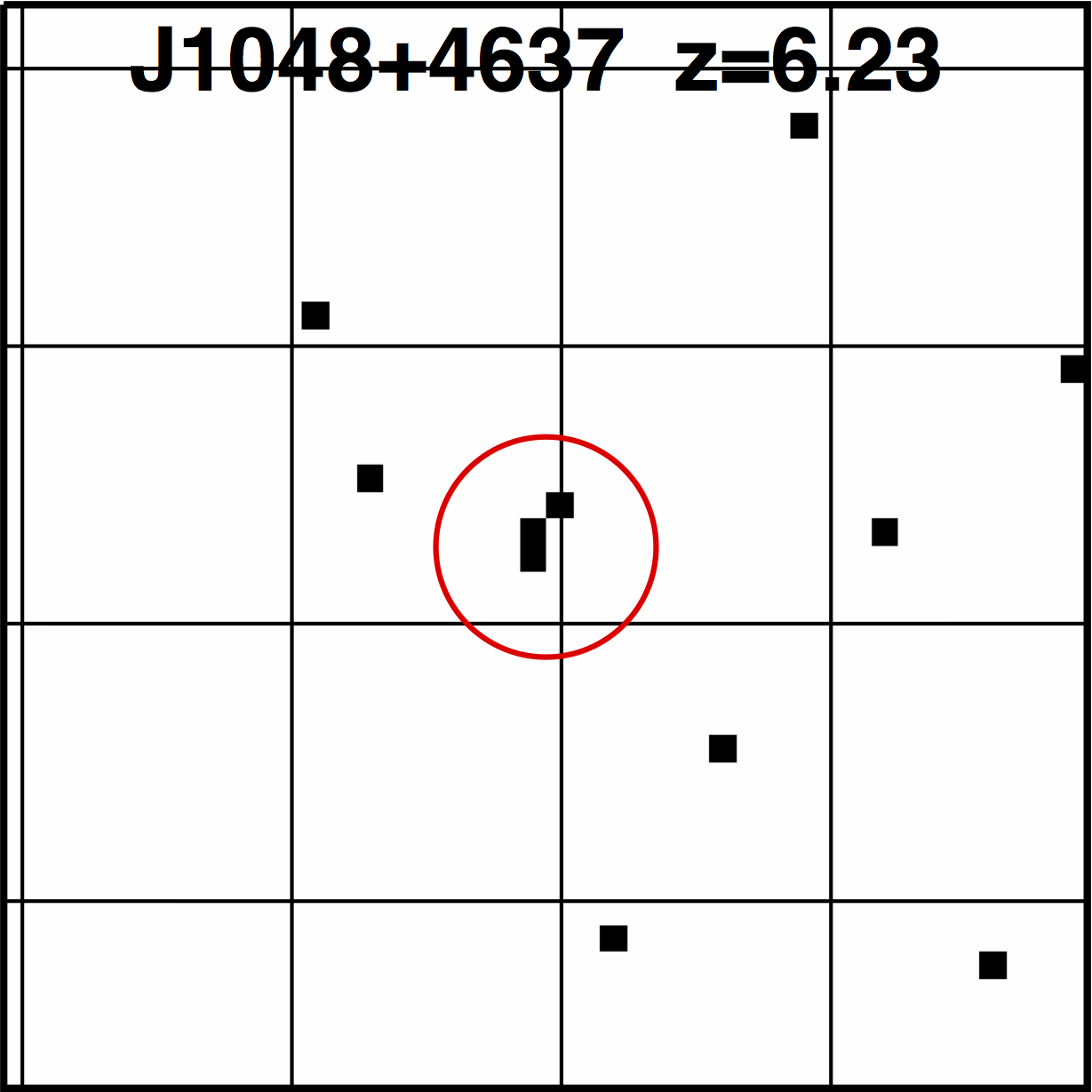} \quad \includegraphics[height=4cm, width=4cm, keepaspectratio]{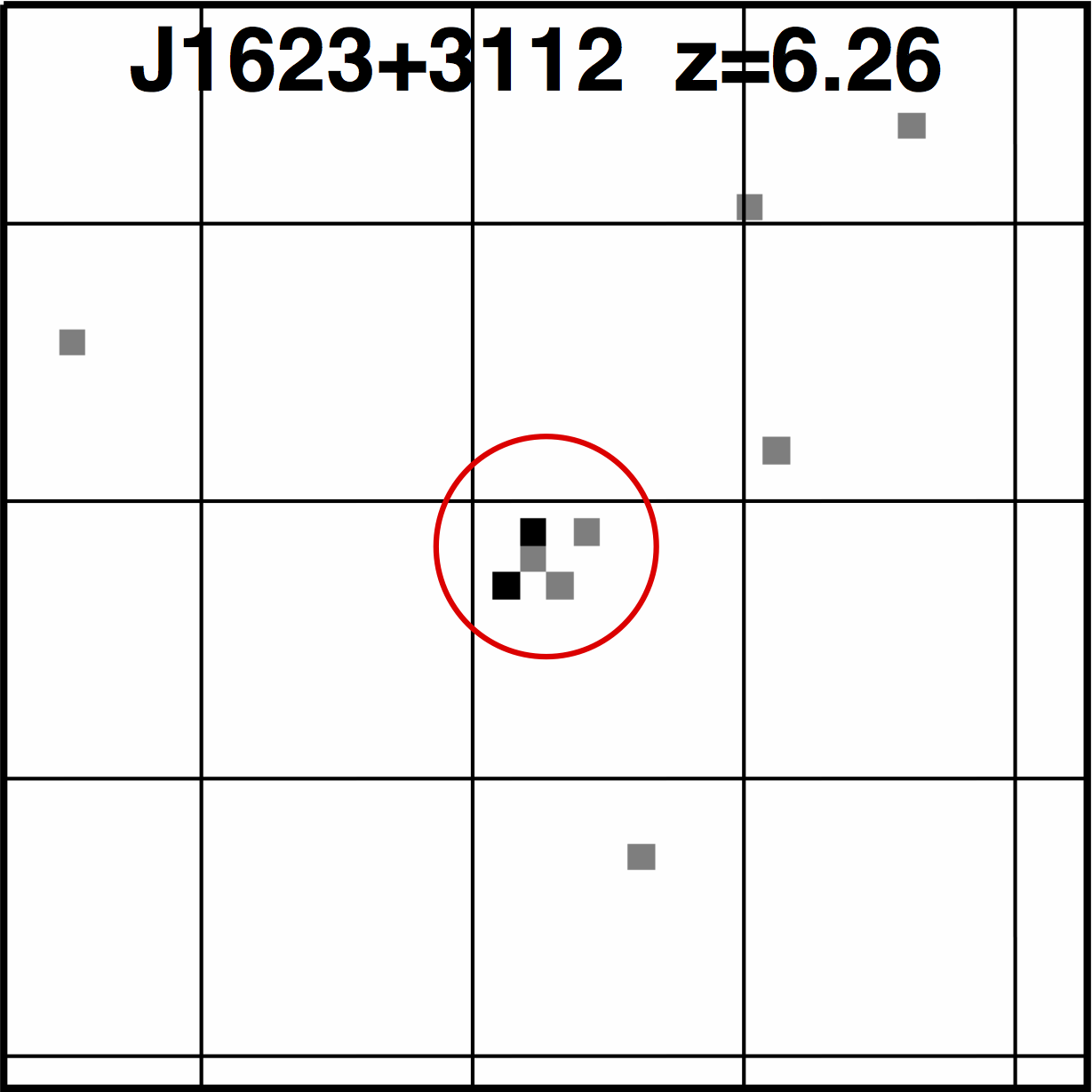}
 \vspace{0.1cm}
 \includegraphics[height=4.005cm, width=4cm]{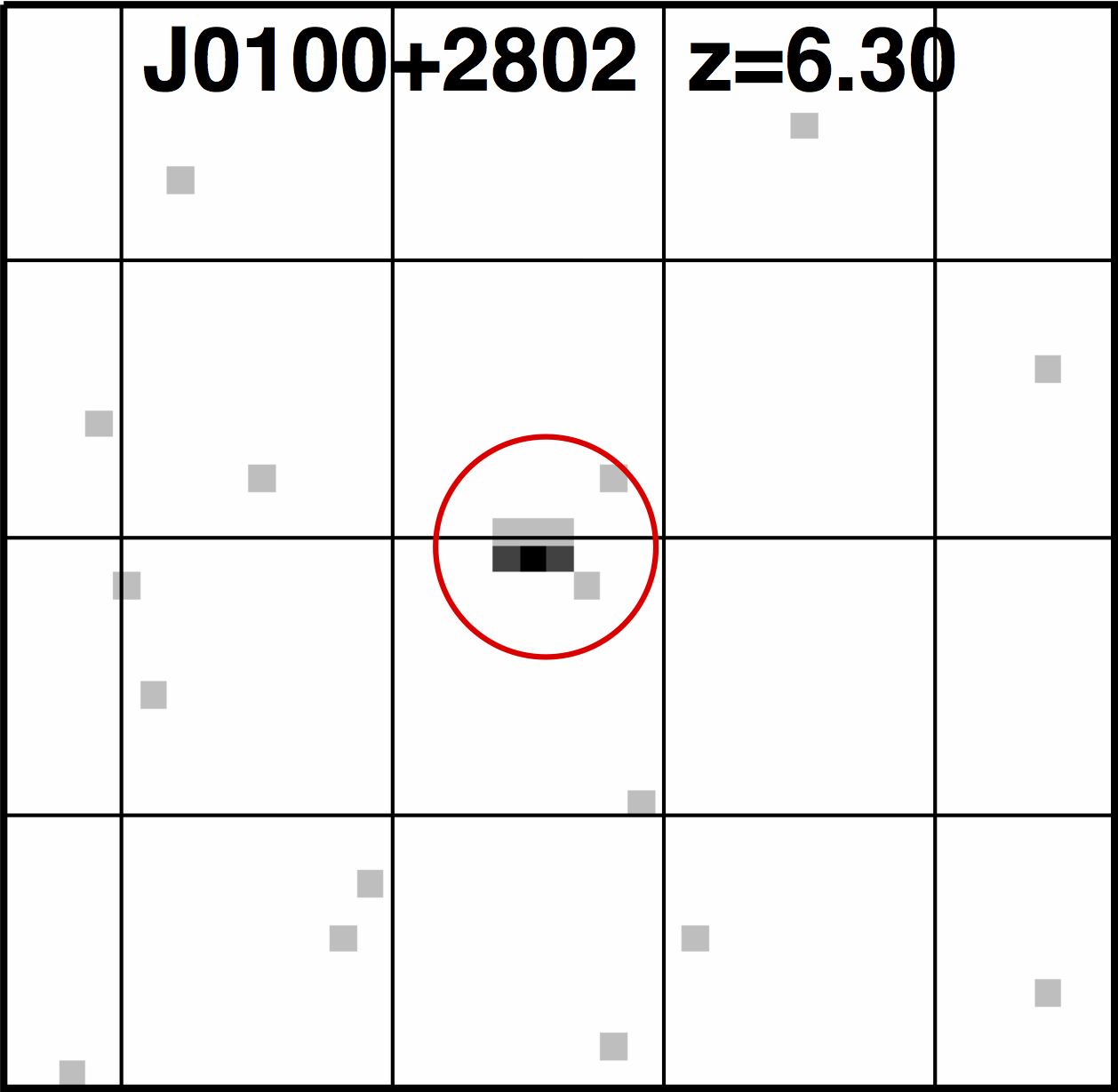} \quad \includegraphics[height=4cm, width=4cm, keepaspectratio]{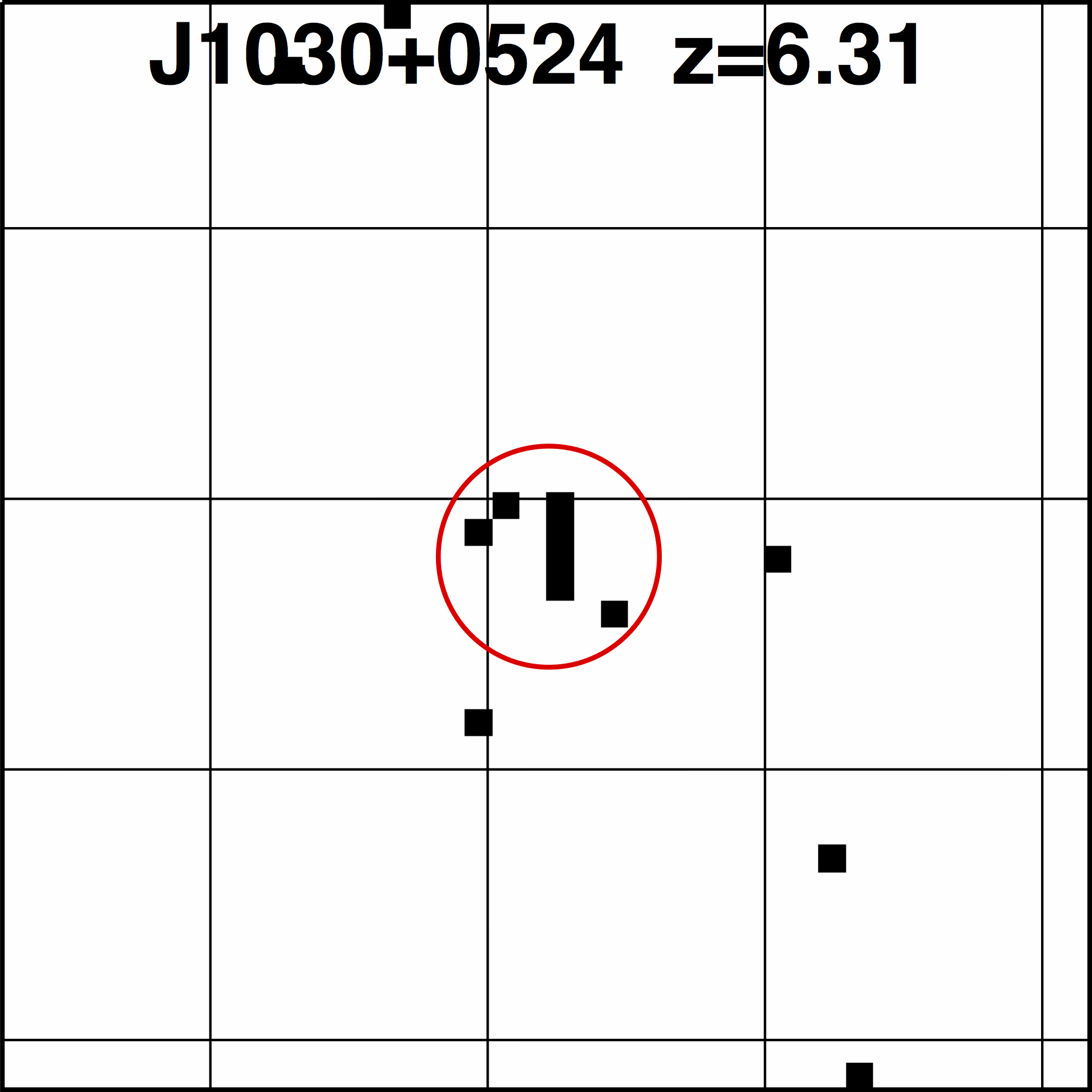}
 \quad \includegraphics[height=4cm, width=4cm, keepaspectratio]{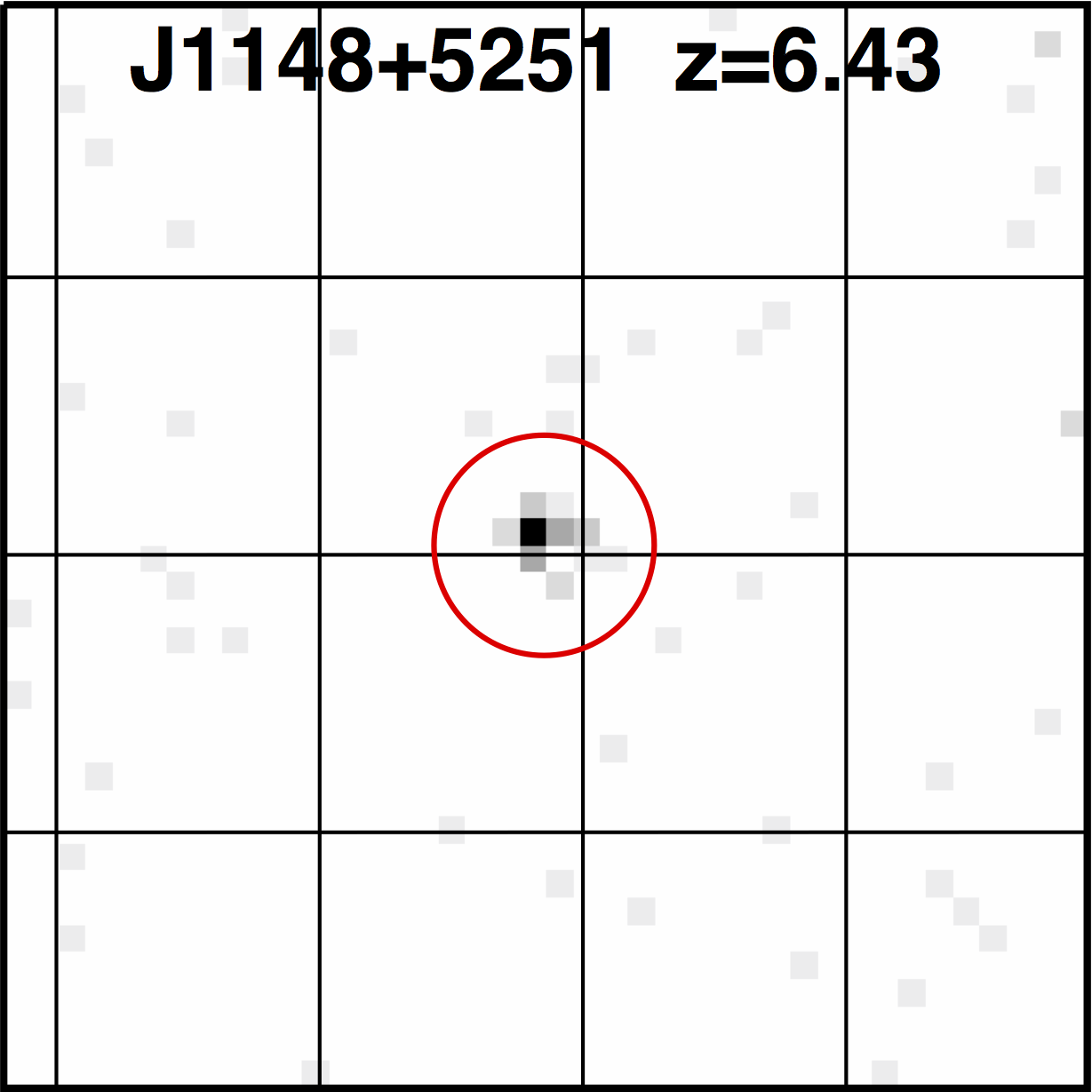} \quad \includegraphics[height=4cm, width=4cm, keepaspectratio]{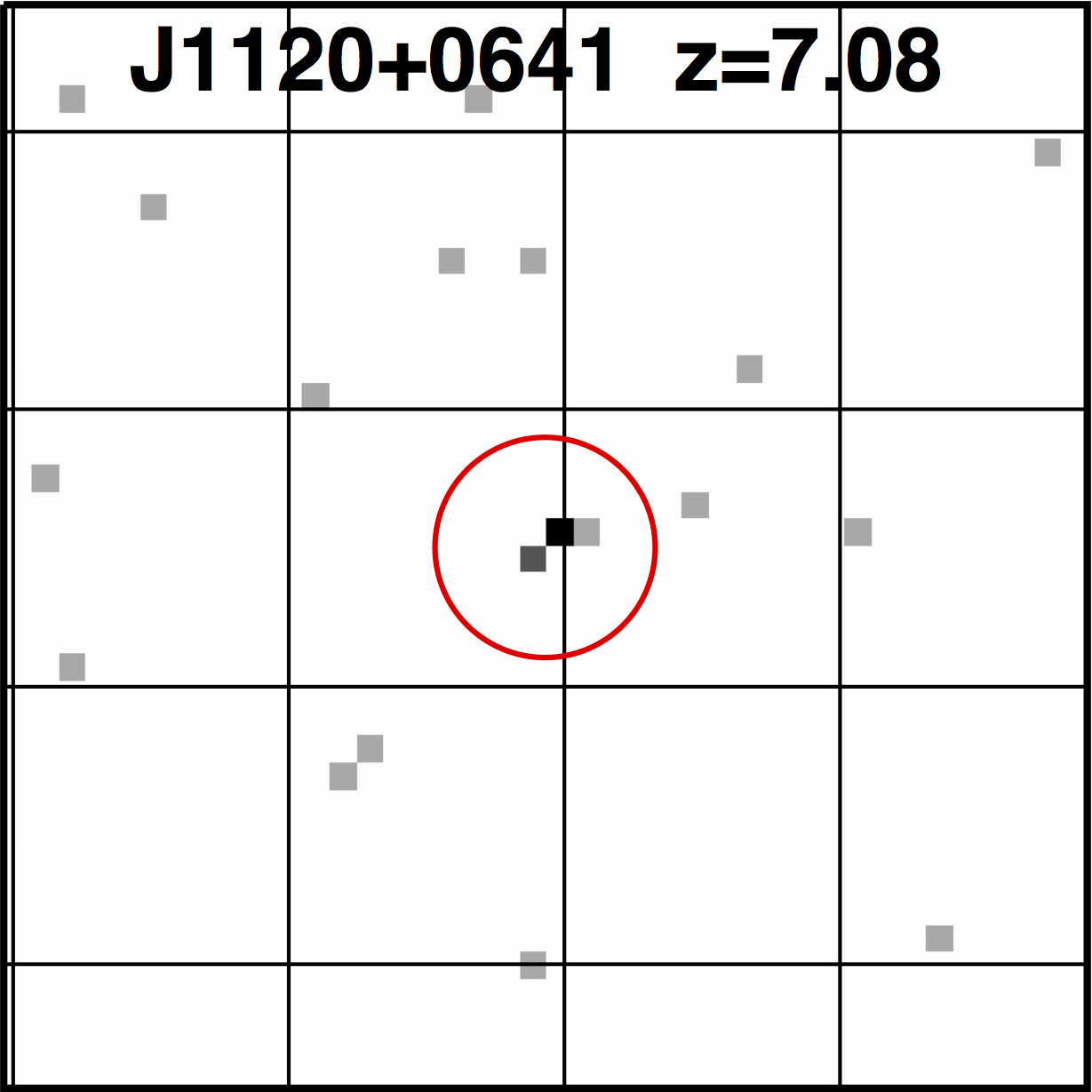}
 \vspace{0.1cm}
 \includegraphics[height=4cm, width=4cm, keepaspectratio]{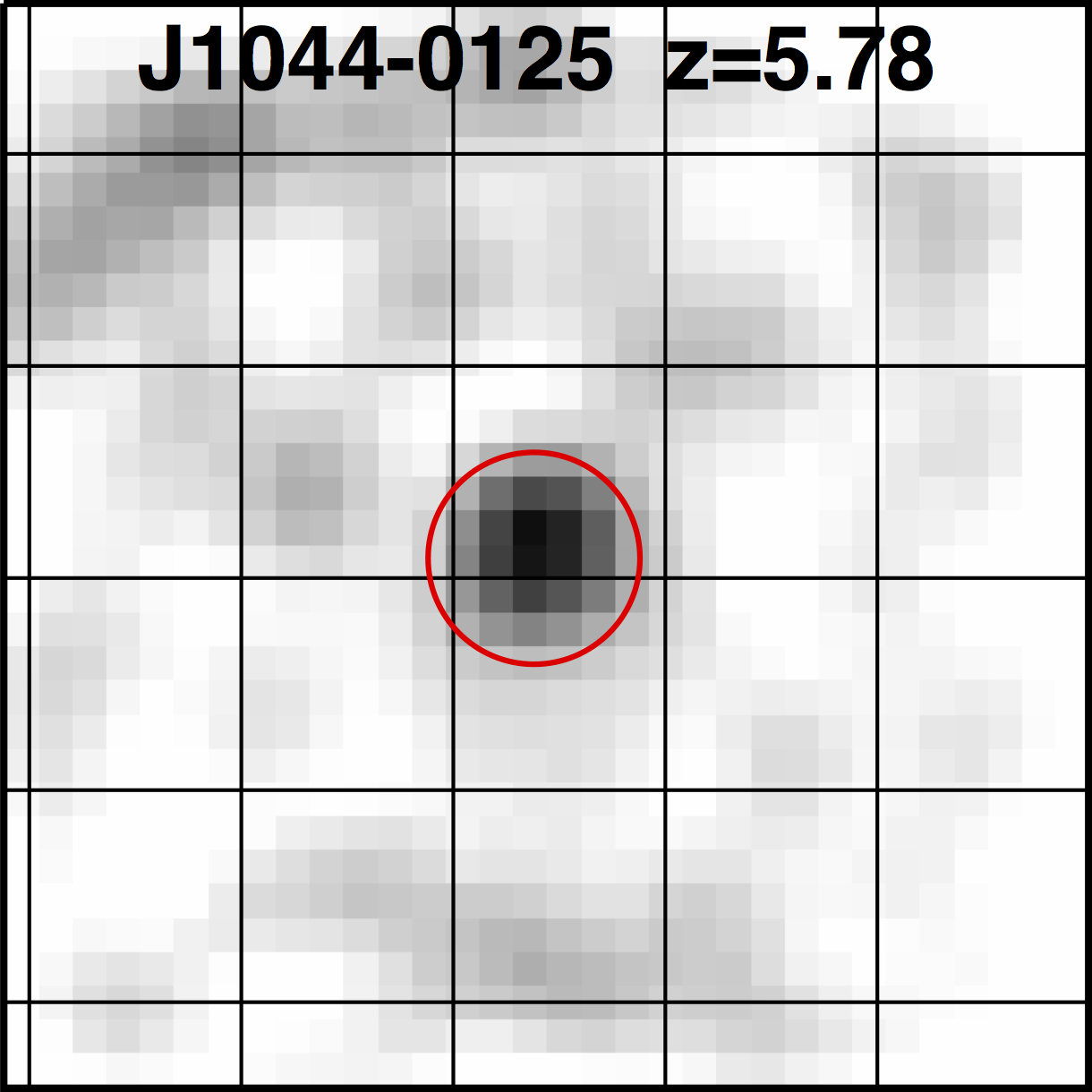} \quad \includegraphics[height=4cm, width=4cm, keepaspectratio]{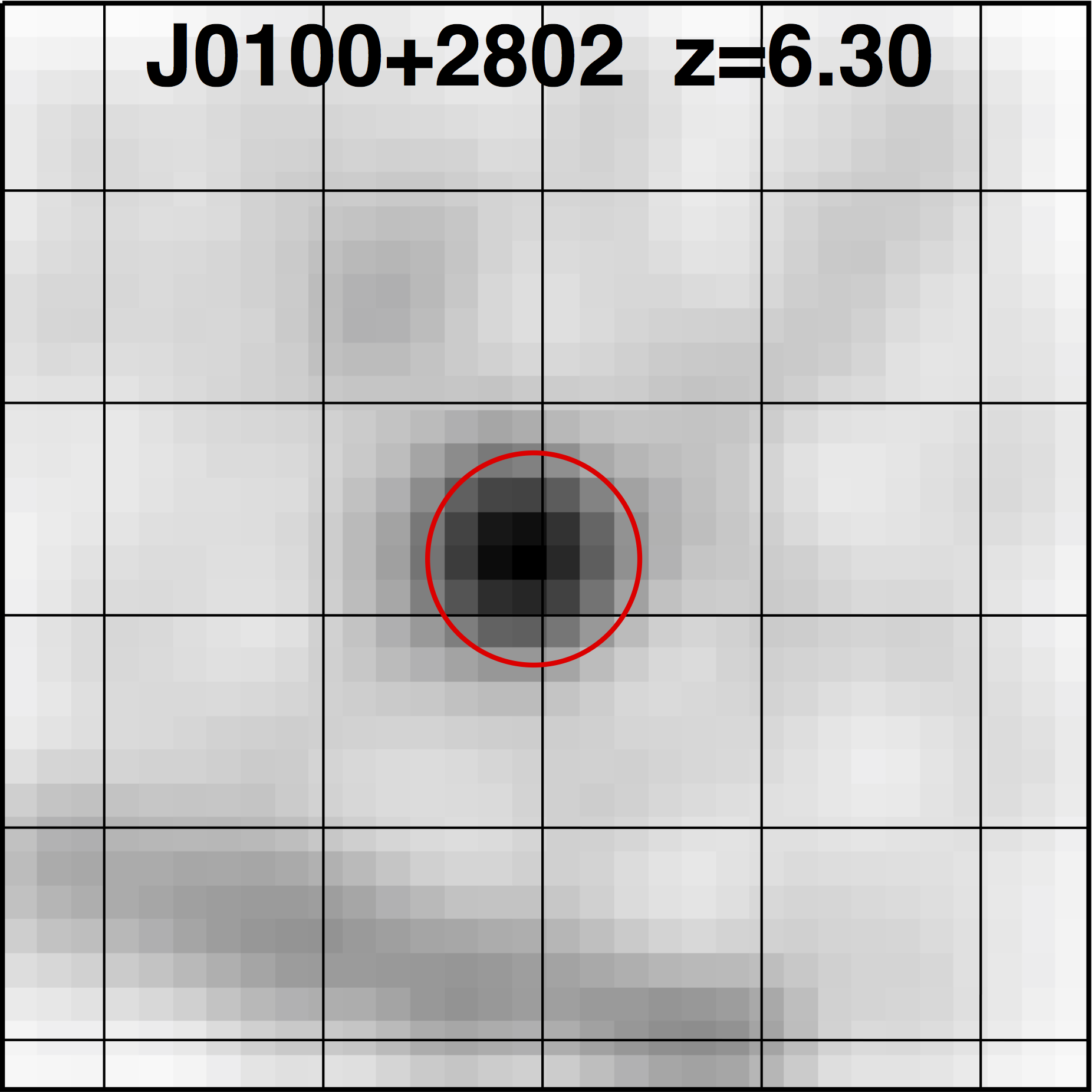}
 \quad \includegraphics[height=4cm, width=4cm, keepaspectratio]{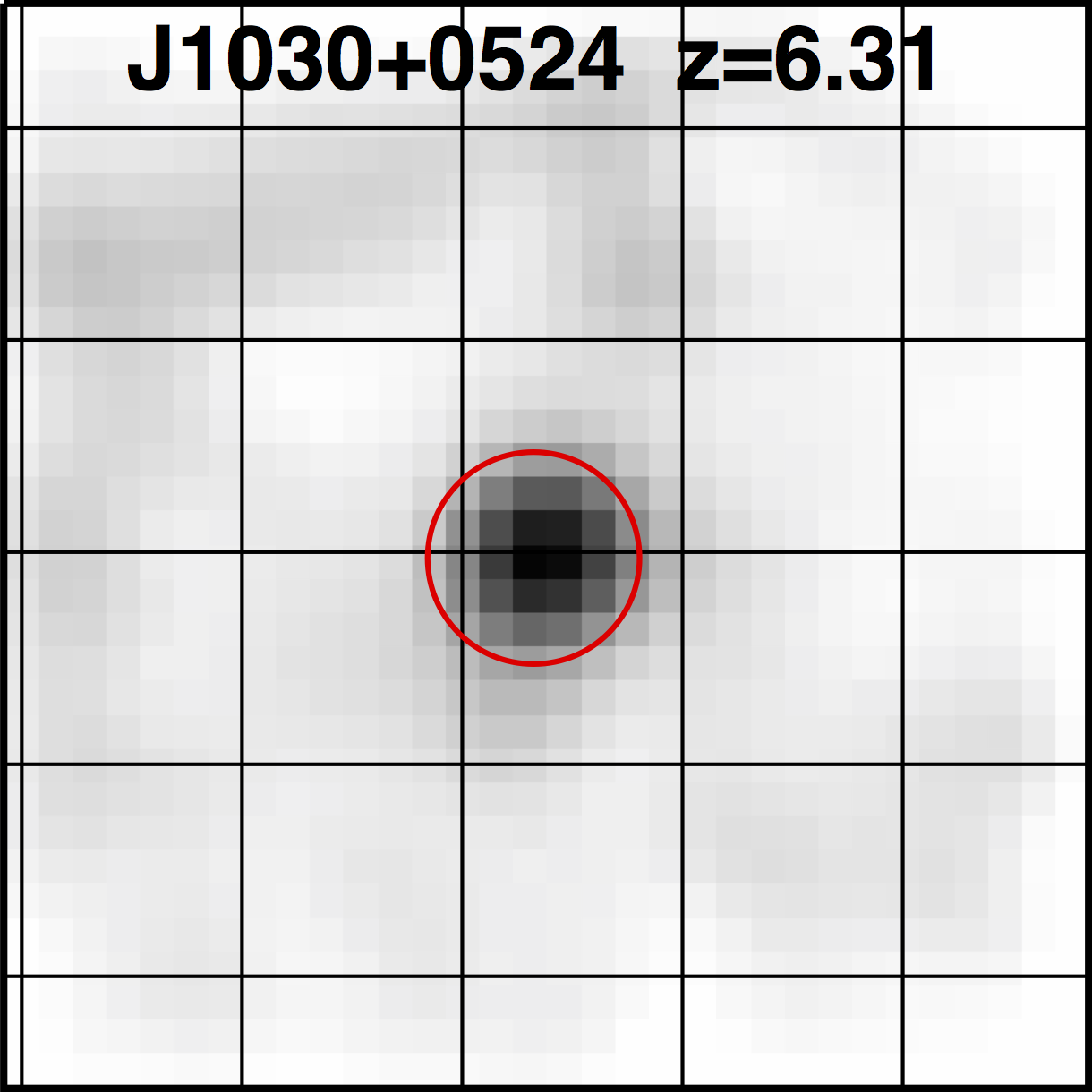} \quad \includegraphics[height=4cm, width=4cm, keepaspectratio]{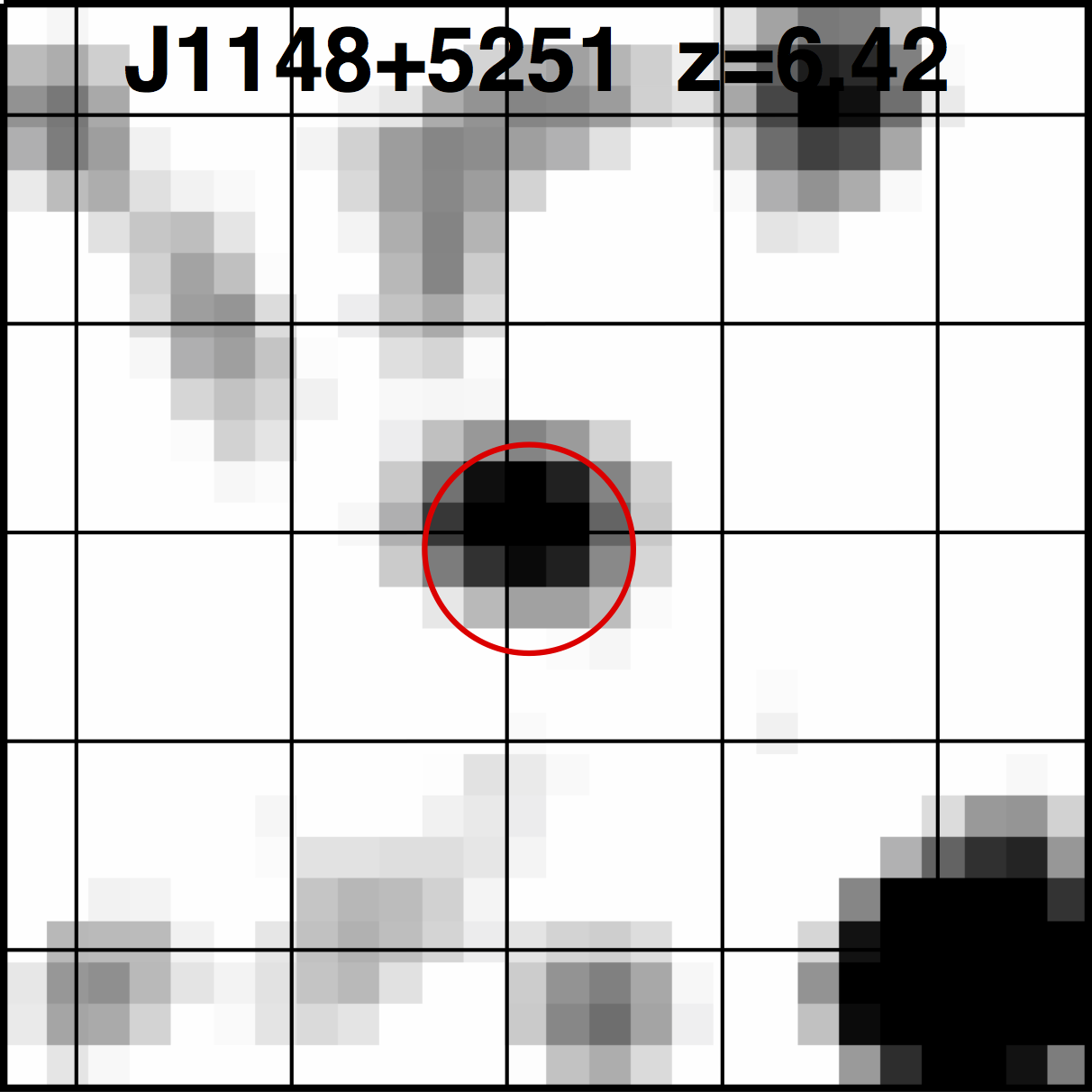} 
 \quad \includegraphics[height=4cm, width=4cm, keepaspectratio]{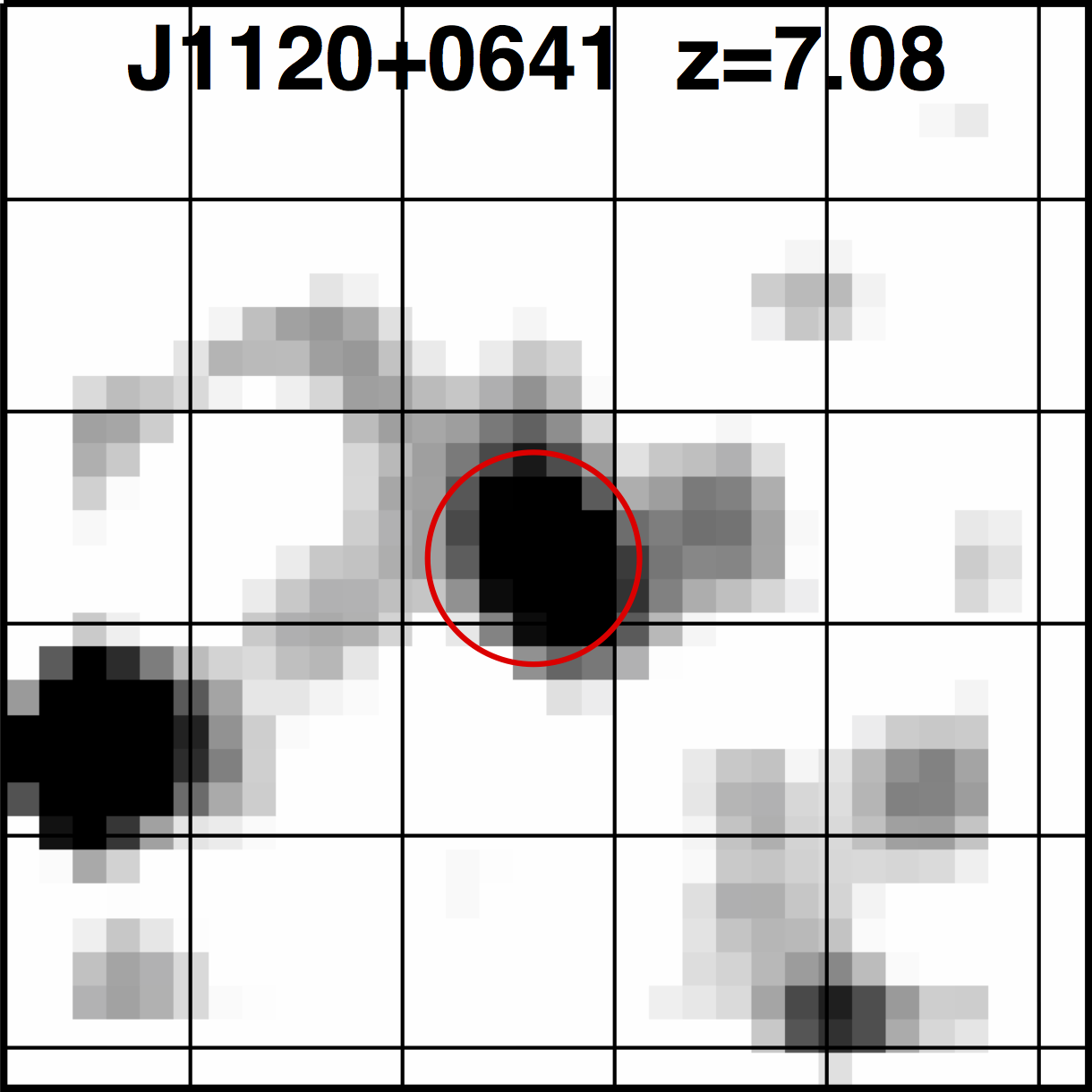} \quad \includegraphics[height=4cm, width=4cm, keepaspectratio]{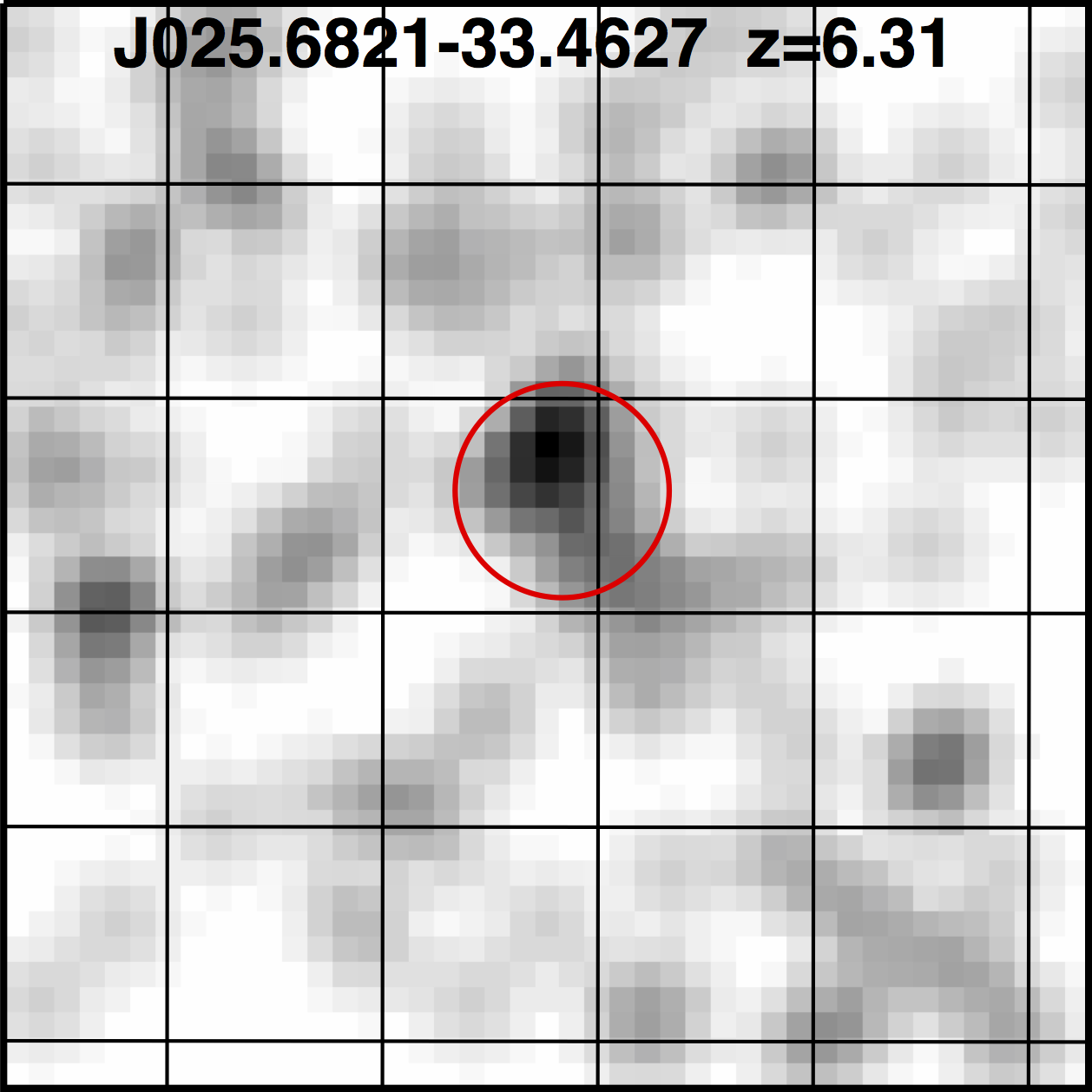}
 \caption{Full-band (0.5-7 keV) cutouts of the 18 detected sources (the first 16 images from \textit{Chandra}, the following five from XMM and the last one from \textit{Swift}). The XMM images are obtained summing data from the three detectors (pn, MOS1, MOS2). Red circles represent our extraction regions (2" and 10" radius for \textit{Chandra} and XMM/\textit{Swift}, respectively). The grid separations are 5" and 20" for \textit{Chandra} and XMM/\textit{Swift} QSOs, respectively. Each panel spans 20"x20" and 100"x100" on the sky for \textit{Chandra} and XMM/\textit{Swift} QSOs, respectively. For clarity the XMM/\textit{Swift} images have been smoothed with a three pixel radius Gaussian function. The XMM cutout of J1148+5251 is shown in the 0.2-12 keV band.}
\end{figure*}

  \begin{table*}
  \centering
  \captionsetup{justification=centering, labelsep = newline}
      \caption[]{X-ray observations log for $z$ > 5.5 quasars with X-ray data}
      \begin{adjustbox}{max width=\textwidth}
         \begin{tabular}{c c c c c c c c c c c}
            \hline
            \hline
            Object & $z$ & R.A. & Decl. & C/X$^a$ & X-ray obs. date C/X & Chip-id C$^b$ & $t_{exp}$ C/X$^c$ & $\theta$ C/X$^d$ & $N_H^g$ & Ref.$^h$\\
            & & & & & & & [ks] & [arcmin] & [10$^{20}$ cm$^{-2}$]\\
            \hline
            NDWFS J142729.7+352209.0 & 5.53 & 14:27:29.70 & +35:22:09.00 & C & 2004 Mar 31 & AI(1) & 4.7 & 7.9 & 1.3 & ...\\
            RD J114816.2+525339.3$\dagger$ & 5.70 & 11:48:16.21 & +52:53:39.30 &  C/X & 2015 Sep 2/2004 Nov 4 & AS(7) & 77.8/3.8 & 1.8/1.5 & 1.5 & 1/...\\   
 	   SDSS J012958.5$-$003539.7$\ddagger$  & 5.78 & 01:29:58.51 & $-$00:35:39.70  & X & 2015 Jan 27 & ... & 1.7 & 15.7 & 3.0 & ...\\   
  	   SDSS J104433.0$-$012502.2$\dagger$ & 5.78 & 10:44:33.04 & $-$01:25:02.20 & X & 2000 May 28 & ... & 31.5 &  0.0 & 4.1 & 2/3\\   
  	   SDSS J083643.8+005453.2 & 5.81 & 08:36:43.86 & +00:54:53.26 & C & 2002 Jan 29 & AS(7) &  5.7  & 0.6 & 4.4 & 2/...\\ 
  	  SDSS J000239.4+255034.9  & 5.82 & 00:02:39.39 & +25:50:34.96 & C & 2005 Jan 24 & AS(7) & 5.9 & 0.6 & 3.2 & 4/...\\ 
 	  SDSS J084035.1+562420.2$\ddagger$ & 5.84 & 08:40:35.10 & +56:24:20.22 & C & 2005 Feb 3 & AS(7) & 15.8. & 0.6 & 4.2 & 4/...\\
  	  SDSS J000552.3$-$000655.8$\diamond$ & 5.85 & 00:05:52.34 & $-$00:06:55.80 & C & 2005 Jul 28 & AS(7) & 16.9 & 0.6 & 3.0 & 4/...\\
	  NDWFS J142516.3+325409.3 & 5.89 & 14:25:16.33 & +32:54:09.54 & C & 2003 Mas 26 &  AI(0) & 4.7 & 3.2 & 1.0 & ...\\
 	  SDSS J133550.8+353315.8$\ddagger$ & 5.90 & 13:35:50.81 & +35:33:15.82 & C & 2008 Mar 10 & AS(7) & 23.5 & 0.3 & 1.0 & ...\\
  	  SDSS J141111.3+121737.3  & 5.90 & 14:11:11.29 & +12:17:37.28 & C & 2005 Mar 17 & AS(7) & 14.3& 0.6 & 1.8 & 4/...\\ 
  	  PSO J328.7339$-$09.5076 & 5.92 & 21:54:56.16  & $-$09:30:27.46 & X & 2004 Nov 1 & ... & 23.5 & 8.2 & 3.8 & ...\\
 	  SDSS J205321.8+004706.8$\ddagger$ & 5.92 & 20:53:21.77 & +00:47:06.80 & C & 2014 Dec 20 & AI(0) & 9.9  & 8.6 & 4.0 & ...\\ 
  	  ULAS J014837.6+060020.1$\dagger$ & 5.98 & 01:48:37.64 & +06:00:20.06 & X & 2002 Jul 14 & ... & 9.3 &  7.6 & 4.7 & ...\\
  	  PSO J007.0273+04.9571$\ddagger$ & 5.99 & 00:28:06.56  & +04:57:25.64 & C & 2001 Dec 7 &  AS(6) & 66.4 & 8.3 & 2.9 & ...\\  
  	  CFHQS J021627.8$-$045534.1 & 6.01 & 02:16:27.81 & $-$04:55:34.10 & X & 2002 Aug 12 & ... & 33.1 & 4.5 & 2.0 & ...\\ 
  	  SDSS J130608.3+035626.4  & 6.02 & 13:06:08.27  & +03:56:26.36 & C & 2004 Dec 11 & AS(7) & 118.2 & 1.0 & 2.0 & 2/...\\
 		"$\; ^e$			& 6.02 & 13:06:08.27 & +03:56:26.36 & C & 2002 Feb 5 & AS(7) &  8.2 & 0.6 & 2.0 & 2/...\\ 
  	   SDSS J163033.9+401209.7 & 6.07 & 16:30:33.90 & +40:12:09.69 & C & 2005 Nov 4 & AS(7) & 27.4 & 0.1 & 1.0 & 4/...\\
  	   SDSS J030331.4$-$001912.9$\diamond$  & 6.08 & 03:03:31.40 & $-$00:19:12.90 & C & 2011 Nov 27 & AS(7) & 1.5 &  4.8 & 6.9 & ...\\ 
  	   SDSS J160253.9+422824.9 & 6.09 & 16:02:53.98 & +42:28:24.94 & C  & 2005 Oct 29 & AS(7) & 13.2 & 0.2 & 1.2 & 4/...\\ 
  	   HSC J221644.5$-$001650.1$\dagger$ & 6.10 & 22:16:44.47 & $-$00:16:50.10 & X & 2011 Dec 7 & ... &  3.7 & 13.9 & 4.9 & ...\\ 
  	   SDSS J104845.1+463718.6$\dagger$ & 6.23 & 10:48:45.07 & +46:37:18.55 & C & 2005 Jan 10 & AS(7) & 15.0 & 0.6 & 1.4 & 4/...\\ 
  	   SDSS J162331.8+311200.5 & 6.26 & 16:23:31.81 & +31:12:00.53 & C & 2004 Dec 29 & AS(7) & 17.2 & 0.6 & 1.8 & 4/...\\ 
  	   SDSS J010013.0+280225.9$\ddagger$ & 6.30 & 01:00:13.02 & +28:02:25.92 & C/X & 2015 Oct 16/2016 Jun 29 & AS(7) & 14.8/46.3 &  0.3/0.0 & 5.8 & 5/...\\  
	   ATLAS J025.6821$-$33.4627 & 6.31 & 01:42:43.73 & $-$33:27:45.47 & S & 2007 Sep 11 \& 2007 Oct 3 & ... & 193.6 $^f$& 4.7 $^f$ & 4.3 & ...\\
  	   SDSS J103027.1+052455.1 & 6.31 & 10:30:27.11 & +05:24:55.06 & C/X & 2002 Jan 29/2003 May 22 & AS(7) & 8.0/51.1 & 0.6/0.0 & 2.6 & 2/6\\  
  	   SDSS J114816.7+525150.4 & 6.42 & 11:48:16.65 & +52:51:50.39 & C/X & 2015 Sep 2/2004 Nov 4 & AS(7) & 77.8/3.8 & 0.3/0.0 & 1.5 & 1/...\\ 
  	   CFHQS J021013.2$-$045620.9 & 6.43 & 02:10:13.19 & $-$04:56:20.90 & X & 2012 Jul 10 & ... & 5.0 & 6.3 & 1.9 & ...\\ 
  	   ULAS J112001.5+064124.3 & 7.08 & 11:20:01.48 & +06:41:24.30 & C/X & 2011 Feb 4/2012 May 23 & AS(7) & 15.8/183.6 & 0.3/0.0 & 5.1 & .../7,8\\
            \hline
         \end{tabular}
        \end{adjustbox}
        \begin{tablenotes}\footnotesize
        		\item Notes - For the XMM exposure time and off-axis angle we provide only the information about the EPIC pn camera.
      		\begin{enumerate}[(a)]
			\item Public data from \textit{Chandra} (C) and/or XMM (X) or \textit{Swift}-XRT (S).
			\item \textit{Chandra} chip identification in which the source is observed. AI stands for ACIS-I (the aimpoint is on the chip ID 3) and AS stands for ACIS-S (the aimpoint is on the chip ID 7).
			\item Exposure time filtered from high-energy time intervals after flare removal.
			\item Off-axis angle of the source.
			\item J1306+0356 has two data-sets taken from two different observations.
			\item This information refers to \textit{Swift}-XRT observations.
			\item Galactic column density calculated using the $nh$ FTOOL (N$_H$ values from \citealt{Kal05}).
			\item References for objects previously published in X-rays: (1) \citet{Gal17}; (2) \citet{Bra02}; (3) \citet{Bra01}; (4) \citet{She06}; (5) \citet{Ai16}; (6) \citet{Far04}; (7) \citet{Mor14}; (8) \citet{Pag14};.
		\end{enumerate}
		\item $\dagger$ BALQs; $\ddagger$ WLQs; $\diamond$ Weak-IR QSOs found in the literature. See \S 5.2.
	 \end{tablenotes}
   \end{table*}   


\section{X-ray analysis of the sample}

In Table 2 we present the number of counts in the total \linebreak (0.5-7.0 keV), soft (0.5-2.0 keV), and hard (2.0-7.0 keV) bands for all the sources; for the undetected QSOs we provide upper limits to the number of counts at the 3$\sigma$ confidence level. For \textit{Chandra} sources, these upper limits were computed using the \textit{srcflux} tool of CIAO, that extracts source and background counts from a circular region, centered at the source position, that contains 90\% of the PSF at 1 keV. For the XMM undetected sources we used the \textit{sosta} command of the XIMAGE software, extracting source and background counts from circular regions with radius r=10" and r=30", respectively. Table 2 also includes the hardness ratio (HR), computed as $HR = \frac{H-S}{H+S}$ where H and S are the net counts in the hard (2-7 keV) and soft (0.5-2.0 keV) bands, respectively. In Figure 3 we report the redshift distribution of the net counts from all sources; upper limits correspond to undetected QSOs. It is evident that the majority of the detected sources have < 30 net counts. \\
For the 12 sources with > 10 counts we attempted an X-ray spectral fit, while we use the tool PIMMS for those QSOs detected with < 10 net counts and those undetected in order to derive the basic X-ray properties.
We "grouped" the spectra ensuring a minimum of one count for each bin, and the best fit was calculated using the Cash statistic\footnote{With a binning of one count for each bin the empty channels are avoided and so the C-stat value is independent of the number of counts. Consequently, the distribution of the C-stat/d.o.f. is centered at $\sim$1. See Appendix A of \citet{Lan13}.},
except for J1306+0356 for which we used a grouping of 20 counts per bin and the $\chi^2$ statistic because of its large number of net counts ($\sim$125). 
We modeled these spectra with an absorbed power-law, using XSPEC v. 12.9 (\citealt{Arn96}). The absorption term takes into account both the Galactic absorption (shown in Table 1) and the source intrinsic obscuration. In the fit we fixed the value of the photon index to $\Gamma=1.9$, which is a typical value found for Type 1 AGN at lower redshift (e.g., \citealt{Pic05}).
We list in Table 3 the basic parameters derived from spectral fits. Errors are reported at 90\% confidence level if not specified otherwise.\\
\begin{figure}
 \centering
 \includegraphics[height=7cm, width=7cm, keepaspectratio]{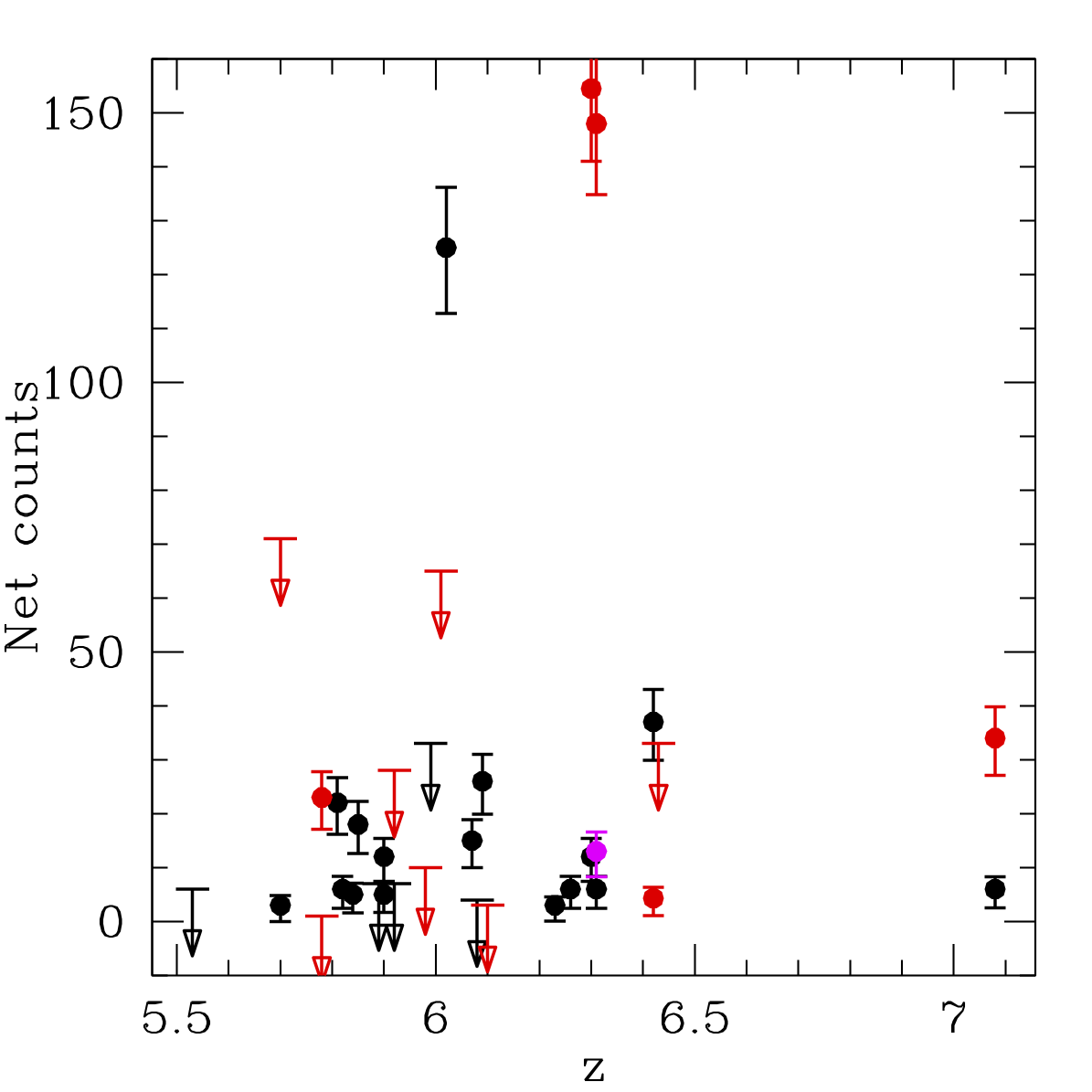}
 \caption{Net counts vs. redshift. Black points represent the 21 QSOs observed with \textit{Chandra}, red points indicate the 12 QSOs observed with XMM (five sources are in common with \textit{Chandra}) and the magenta point represents the QSO observed with \textit{Swift}-XRT. Detected sources are shown as full dots, while arrows represent 3$\sigma$ upper limits.}
\end{figure}
We also fit the five spectra of the sources with highest counting statistics (> 30 net counts) using the same model described above but with $\Gamma$ free to vary. We present these results in the next sub-section.

\begin{table*}[!ht]
  \begin{center}
  \captionsetup{justification=centering, labelsep = newline}
      \caption[]{X-ray counts and hardness ratio}
      \begin{adjustbox}{center, max width=\textwidth,}
         \begin{tabular}{c c c c c}
            \hline
            \hline
            \multirow{2}{*}{Object} & \multicolumn{3}{c}{X-ray counts \textit{Chandra/XMM$^a$}} & \multirow{2}{*}{HR$^b$} \\
            \hhline{~---}
             & 0.5$-$7.0 keV & 0.5$-$2.0 keV & 2.0$-$7.0 keV & \\
             \hline \rule[0.7mm]{0mm}{3.5mm}
           NDWFS J1427+3522 & < 6.4 & < 5.0 & < 1.4 & $-$ \\
           \rule[0.7mm]{0mm}{3.5mm}
           RD J1148+5253 & 3.3$_{-1.8}^{+3.0}$ & 0.9$_{-0.8}^{+2.3}$ & 2.4$_{-1.5}^{+2.8}$ & $-$0.45$_{-0.65}^{+0.65}$\\
           \rule[0.7mm]{0mm}{3.5mm}
           " $^c$ & < 71.3 & < 57.9 & < 13.4 & $-$ \\
            \rule[0.7mm]{0mm}{3.5mm}
           SDSS J0129$-$0035 $^c$ & < 1.3 & < 1.1 & < 0.2 & $-$ \\
           \rule[0.7mm]{0mm}{3.5mm}
           SDSS J1044$-$0125 $^c$ & 22.9$_{-4.8}^{+5.9}$ & 19.2$_{-4.4}^{+5.5}$ & 3.1$_{-1.7}^{+3.0}$ & $-$0.72$_{-0.16}^{+0.21}$ \\   
           \rule[0.7mm]{0mm}{3.5mm}
  	   SDSS J0836+0054 & 21.9$_{-4.7}^{+5.8}$ & 17.8$_{-4.2}^{+5.3}$ &  3.0$_{-1.7}^{+2.9}$  & $-$0.71$_{-0.17}^{+0.22}$ \\ 
	   \rule[0.7mm]{0mm}{3.5mm}
  	  SDSS J0002+2550  & 5.8$_{-2.4}^{+3.6}$ & 5.8$_{-2.4}^{+3.6}$ & < 3.0 & < $-$0.32 \\ 
	  \rule[0.7mm]{0mm}{3.5mm}
 	  SDSS J0840+5624 & 4.8$_{-2.1}^{+3.4}$ & 3.8$_{-1.9}^{+3.1}$ & 1.0$_{-0.9}^{+2.3}$ & $-$0.58$_{-0.46}^{+0.54}$ \\
	  \rule[0.7mm]{0mm}{3.5mm}
  	  SDSS J0005$-$0006 & 18.4$_{-4.3}^{+5.4}$ & 16.6$_{-4.0}^{+5.2}$ & 1.6$_{-1.2}^{+2.5}$ & $-$0.82$_{-0.14}^{+0.22}$  \\
	  \rule[0.7mm]{0mm}{3.5mm}
	  NDWFS J1425+3254 & < 6.7 & < 5.3 & < 1.4 & $-$ \\
	  \rule[0.7mm]{0mm}{3.5mm}
 	  SDSS J1335+3533 & 4.6$_{-2.1}^{+3.3}$ & 3.8$_{-1.9}^{+3.1}$ & 0.8$_{-0.7}^{+2.2}$ & $-$0.56$_{-0.52}^{+0.62}$ \\
	  \rule[0.7mm]{0mm}{3.5mm}
  	  SDSS J1411+1217 & 11.9$_{-3.4}^{+4.6}$ & 9.9$_{-3.1}^{+4.3}$ & 2.0$_{-1.3}^{+2.7}$ & $-$0.66$_{-0.24}^{+0.32}$ \\ 
	  \rule[0.7mm]{0mm}{3.5mm}
	  PSO J328.7339$-$09.5076 $^c$ & < 28.2 & < 22.9 & < 5.3 & $-$ \\
	  \rule[0.7mm]{0mm}{3.5mm}
	  SDSS J2053+0047 & < 6.6 & < 4.7 & < 1.9 & $-$ \\
	  \rule[0.7mm]{0mm}{3.5mm}
	  ULAS J0148+0600 $^c$ & < 10.1 & < 8.2 & < 1.9 & $-$ \\
	  \rule[0.7mm]{0mm}{3.5mm}
	  PSO J007.0273+04.9571 & < 33.2 & < 28.2 & < 5.0 & $-$ \\
	  \rule[0.7mm]{0mm}{3.5mm}
	  CFHQS J0216$-$0455 $^c$ & < 65.2 & < 53.0 & < 12.2 & $-$ \\
	   \rule[0.7mm]{0mm}{3.5mm}
  	  SDSS J1306+0356  & 125.4$_{-11.2}^{+12.2}$ & 87.3$_{-9.3}^{+10.4}$ & 38.1$_{-6.2}^{+7.2}$ & $-$0.39$_{-0.09}^{+0.09}$\\
	  \rule[0.7mm]{0mm}{3.5mm}
   	   SDSS J1630+4012 & 15.3$_{-3.9}^{+5.0}$ & 10.8$_{-3.2}^{+4.4}$ & 4.5$_{-2.1}^{+3.3}$ & $-$0.41$_{-0.27}^{+0.28}$\\
	   \rule[0.7mm]{0mm}{3.5mm}
	   SDSS J0303$-$0019 & < 3.8 & < 3.0 & < 0.8 & $-$ \\
	   \rule[0.7mm]{0mm}{3.5mm}
  	   SDSS J1602+4228 & 25.7$_{-5.0}^{+6.1}$ & 21.8$_{-4.6}^{+5.7}$ & 3.8$_{-1.9}^{+3.1}$ & $-$0.70$_{-0.15}^{+0.19}$\\ 
	    \rule[0.7mm]{0mm}{3.5mm}
	   HSC J2216$-$0016 $^c$ & < 3.0 & < 2.4 & < 0.6 & $-$\\
	   \rule[0.7mm]{0mm}{3.5mm}
    	   SDSS J1048+4637 & 2.9$_{-1.6}^{+2.9}$ & 2.9$_{-1.6}^{+2.9}$ & < 3.0 & < 0.02 \\ 
	   \rule[0.7mm]{0mm}{3.5mm}
  	   SDSS J1623+3112 & 6.0$_{-2.4}^{+3.6}$ & 4.0$_{-1.9}^{+3.2}$ & 2.0$_{-1.3}^{+2.6}$ & $-$0.33$_{-0.48}^{+0.47}$ \\ 
	   \rule[0.7mm]{0mm}{3.5mm}
  	   SDSS J0100+2802 & 12.0$_{-3.4}^{+4.6}$ & 10.7$_{-3.2}^{+4.4}$ & 0.7$_{-0.6}^{+2.2}$ &  $-$0.88$_{-0.11}^{+0.31}$ \\  
	   \rule[0.7mm]{0mm}{3.5mm}
	   " $^c$ & 154.5$_{-12.4}^{+13.5}$ & 127.9$_{-11.3}^{+12.3}$ & 25.8$_{-5.1}^{+6.1}$ & $-$0.66$_{-0.06}^{+0.07}$\\
	   \rule[0.7mm]{0mm}{3.5mm}
	   ATLAS J025.6821$-$33.4627 $^d$ & 13.0$_{-3.6}^{+4.7}$ & 10.4$_{-3.2}^{+4.3}$ & 1.6$_{-1.2}^{+2.5}$ & $-$0.73$_{-0.22}^{+0.31}$\\
	   \rule[0.7mm]{0mm}{3.5mm}
  	   SDSS J1030+0524 & 6.0$_{-2.4}^{+3.6}$ & 6.0$_{-2.4}^{+3.6}$ & < 3.0 & < $-$0.33 \\	   
	   \rule[0.7mm]{0mm}{3.5mm}
	   " $^c$ & 148.0$_{-12.2}^{+13.2}$ & 128.8$_{-11.3}^{+12.3}$ & 19.0$_{-4.3}^{+5.4}$ & $-$0.74$_{-0.06}^{+0.06}$ \\
	   \rule[0.7mm]{0mm}{3.5mm}
  	   SDSS J1148+5251 & 36.8$_{-6.0}^{+7.1}$ & 25.9$_{-5.1}^{+6.2}$ & 10.9$_{-3.3}^{+4.4}$ & $-$0.41$_{-0.17}^{+0.17}$ \\ 
	   \rule[0.7mm]{0mm}{3.5mm}
	   " $^c$ & 4.3$_{-2.0}^{+3.2}$ & 3.1$_{-1.7}^{+3.0}$ & 1.2$_{-1.0}^{+2.4}$ & $-$0.44$_{-0.57}^{+0.57}$\\
	   \rule[0.7mm]{0mm}{3.5mm}
	   CFHQS J0210$-$0456 $^c$ & < 32.8 & < 26.7 & < 6.1 & $-$ \\
	    \rule[0.7mm]{0mm}{3.5mm}
   	   ULAS J1120+0641 & 5.7$_{-2.3}^{+3.5}$ & 4.0$_{-1.9}^{+3.2}$ & 1.7$_{-1.2}^{+2.6}$ & $-$0.40$_{-0.48}^{+0.49}$ \\
	   \rule[0.7mm]{0mm}{3.5mm}
	   " $^c$ & 34.0$_{-5.8}^{+6.9}$ & 30.7$_{-5.5}^{+6.6}$ & 3.1$_{-1.7}^{+3.0}$ & $-$0.82$_{-0.10}^{+0.15}$ \\[2pt]
            \hline
         \end{tabular}
        \end{adjustbox}
        \begin{tablenotes}\footnotesize
        \item	\begin{enumerate}[(a)]
			\item Errors on the X-ray counts were computed according to Table 1 and 2 of \citet{Geh86} and correspond to the 1$\sigma$ level in Gaussian statistics.
			The upper limits are at the 3$\sigma$ confidence level.
			\item The hardness ratio is defined as $HR = \frac{H-S}{H+S}$ where H and S are the counts in the hard (2.0-7.0 keV) and soft (0.5-2.0 keV) bands.
			We calculated errors at the 1$\sigma$ level for the hardness ratio following the method described in \S 1.7.3 of Lyons (1991).
			\item Sources observed by XMM for which we provide EPIC pn information.
			\item Source observed by \textit{Swift}-XRT.
		\end{enumerate}
	 \end{tablenotes}
	 \end{center}
   \end{table*}


\subsection{Analysis of the five QSOs with the best photon statistics}

In this section we show the results obtained from our analysis of the five quasars with the best counting statistics (> 30 net counts): J0100+2802, J1030+0524, J1120+0641, J1148+5253 and J1306+0356. In all the fit models we included a Galactic-absorption component, which was kept fixed during the fit.\\
 
\textit{SDSS J1306+0356} ($z$ = 6.02). This is the only quasar detected by \textit{Chandra} with more than 100 net counts in the \linebreak 0.5-7.0~keV band. The target was observed in two different periods and has two different data-sets. In order to improve the fit quality we combined the two data-sets obtaining a spectrum with $\sim$125 net counts. In the fit we used a grouping of 20 counts per bin in order to use the $\chi^2$ statistic and we were able to fit its spectrum with a model in which the photon index $\Gamma$ was left free to vary. We fit the spectrum with a power-law model at the redshift of the quasar. The spectrum and its best-fit model and residuals are shown in Figure 4 (a). Throughout the paper, residuals are in terms of sigmas with error bars of size one. In the case of the Cash statistic, they are defined as the (data$-$model)/error, where error is calculated as the square root of the model predicted number of counts. The best fit photon index is $\Gamma = 1.72_{-0.52}^{+0.53}$ with $\chi^2 = 2.2$ for 3 degrees of freedom. Such a value of $\Gamma$ is consistent with the others found for luminous AGN at lower redshift (1 $\le$ $z$ $\le$ 5.5; e.g., \citealt{Vig05}; \citealt{She06}; \citealt{Jus07}).
We then added an absorption component at the redshift of the quasar and obtained an upper limit of $N_H < 2.2 \cdot 10^{23}$ cm$^{-2}$.

\textit{SDSS J1030+0524} ($z$ = 6.31). This quasar was observed by both \textit{Chandra} and XMM. The short \textit{Chandra} exposure detected this source with $\sim$6 net counts, while the much longer XMM observation (see Table 1) detected this source with $\sim$148 net counts. We used a grouping of 1 count for each bin for all spectra of the three cameras and we fit the three EPIC spectra (pn, MOS1 and MOS2) with a power-law model and an intrinsic-absorption component at the redshift of the quasar. The spectrum and its best-fit model and residuals are shown in Figure 4 (b). The best-fit photon index is $\Gamma = 2.39_{-0.46}^{+0.55}$ with C-stat = 21.6 for 18 degrees of freedom. This value of $\Gamma$ is consistent with the one found by Farrah et al. (2004; $\Gamma = 2.27_{-0.31}^{+0.31}$). We also found an upper limit to the column density $N_H < 1.9 \cdot 10^{23}$ cm$^{-2}$.

\textit{SDSS J1148+5251} ($z$ = 6.42). This quasar was observed by both observatories; \textit{Chandra} detected this source with $\sim$37 net counts thus allowing us to fit its data. 
We used a grouping of 1 count for each bin, and
we fit the spectrum with a simple power-law model. The spectrum and its best-fit model and residuals are shown in Figure 4 (c). The best-fit photon index is $\Gamma = 1.59_{-0.57}^{+0.61}$ with C-stat = 20.9 for 33 degrees of freedom. This value of $\Gamma$ is consistent with the one found by Gallerani et al. (2017; $\Gamma = 1.6_{-0.49}^{+0.49}$). 

\textit{SDSS J0100+2802} ($z$ = 6.30). This is the latest quasar observed by both \textit{Chandra} and XMM. The \textit{Chandra} exposure detected this source with $\sim$12 net counts, while a total of $\sim$155 net counts were collected by XMM. Fitting a power-law to the \textit{Chandra} spectrum, we obtained $\Gamma = 3.0_{-0.8}^{+1.2}$ which is consistent with the one found by Ai et al. (2016; $\Gamma = 3.03_{-0.70}^{+0.78}$), but this measurement is uncertain with very large errors. 
For the XMM spectrum, we used a grouping of 1 count for each bin for all spectra of the three cameras and we fit the three EPIC spectra (pn, MOS1 and MOS2) with a power-law model and an intrinsic-absorption component at the redshift of the quasar. The spectrum and its best-fit model and residuals are shown in Figure 4 (d). The best-fit photon index is $\Gamma = 2.33_{-0.29}^{+0.32}$ with C-stat = 233.5 for 254 degrees of freedom. This value of $\Gamma$ is consistent with the one found by Ai et al. (2016) and the one we derived from the \textit{Chandra} analysis, but is less uncertain. We also found an upper limit to the column density $N_H < 2.1 \cdot 10^{23}$ cm$^{-2}$.

\textit{ULAS J1120+0641} ($z$ = 7.08). This is another quasar observed by both \textit{Chandra} and XMM (which observed it in three different orbits).
We summed together the six MOS and the three pn spectra and then we summed the two combined MOS, so as to increase the fit quality, and used a binning of 1 count per bin.
\textit{Chandra} detected this source with $\sim$6 net counts and we were not able to fit its data, while XMM detected this source with $\sim$34 net counts. We fit the two EPIC spectra with a power-law model and compared our result with those available in the literature (Moretti et al. 2014; Page et al. 2014). The spectrum and its best-fit model and residuals are shown in Figure 4 (e). The best-fit photon index is $\Gamma = 2.24_{-0.48}^{+0.55}$ with C-stat = 391.1 for 364 degrees of freedom. Such a value of $\Gamma$ is half way between those found by Page et al. (2014; $\Gamma = 2.64_{-0.54}^{+0.61}$) and Moretti et al. (2014; $\Gamma = 1.98_{-0.43}^{+0.46}$) and consistent with both of them within the errors.
\begin{figure*}[!h]
\centering
\includegraphics[height=6.2cm, width=8.9cm]{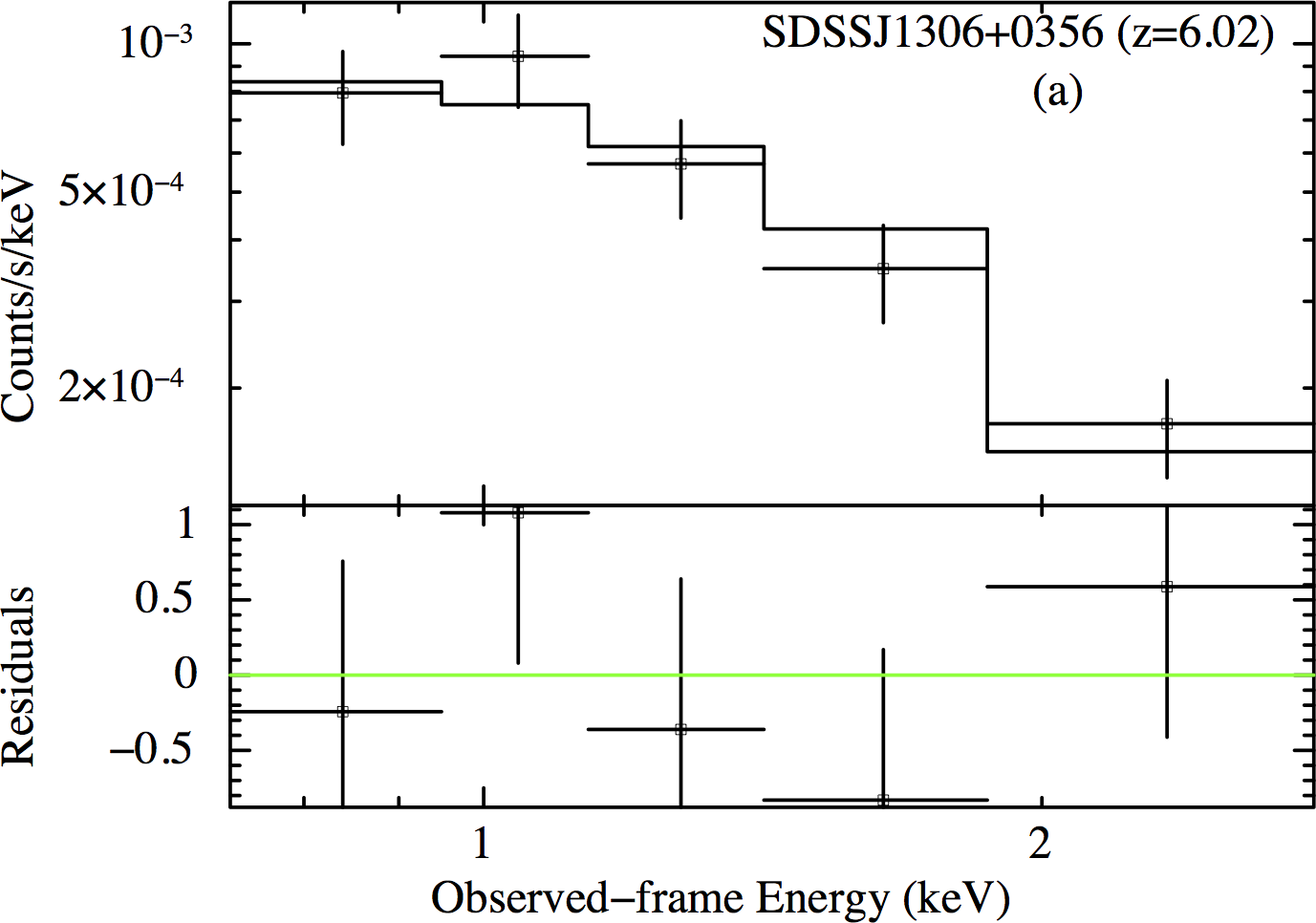} \quad \includegraphics[height=6.2cm, width=8.9cm]{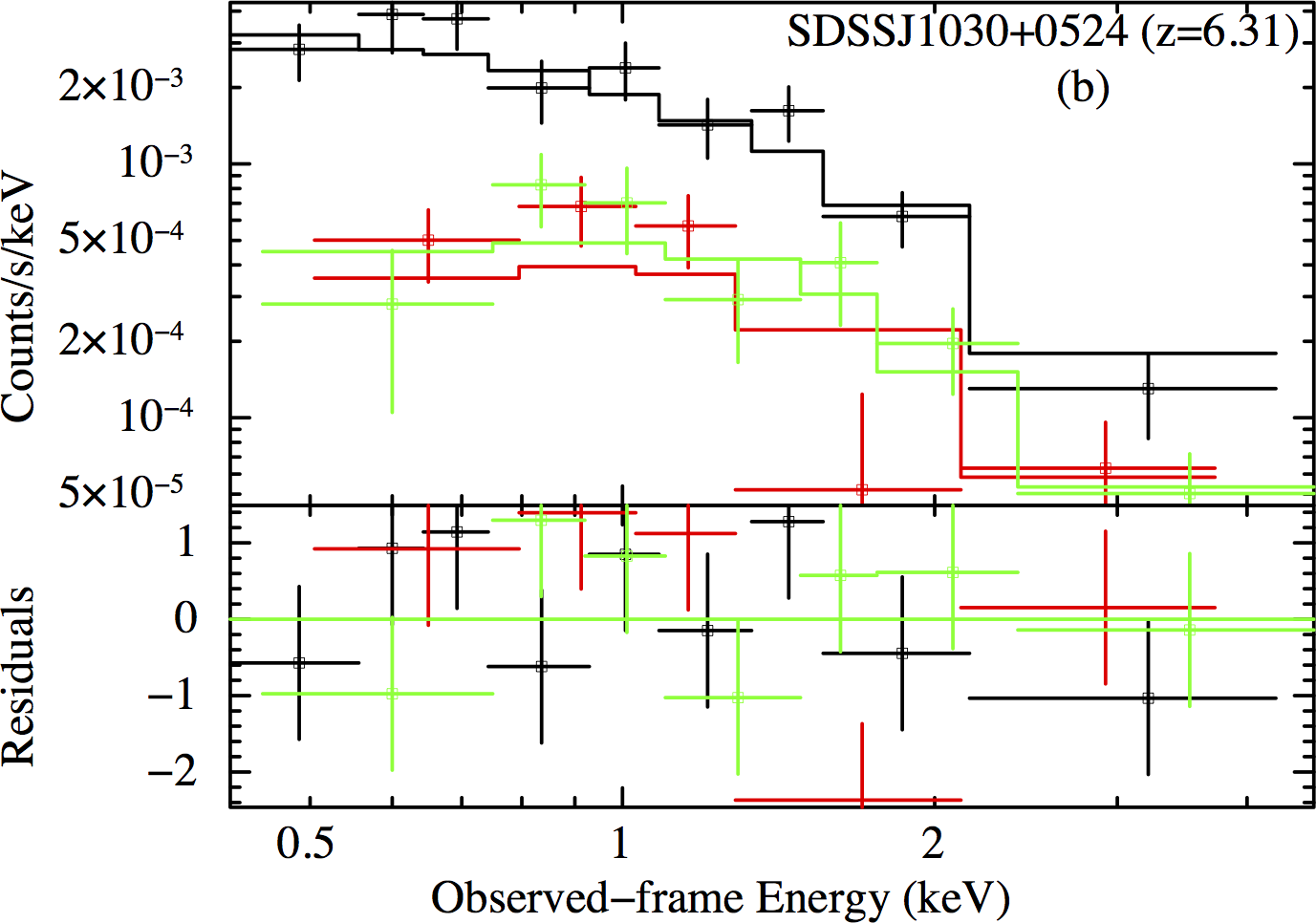} 
\par \vspace{0.5cm}
 \includegraphics[height=6.2cm, width=8.9cm]{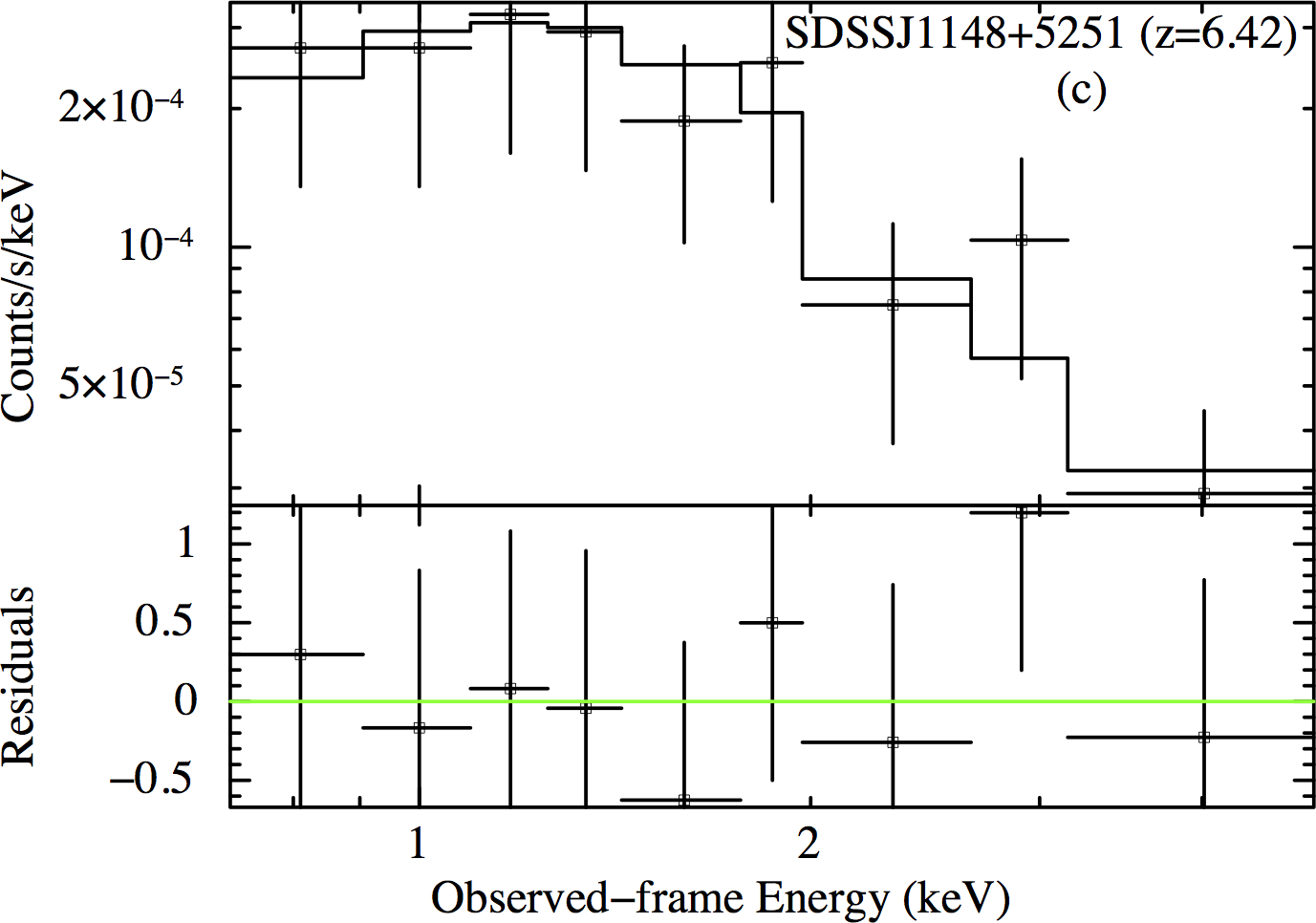} \quad \includegraphics[height=6.2cm, width=8.9cm]{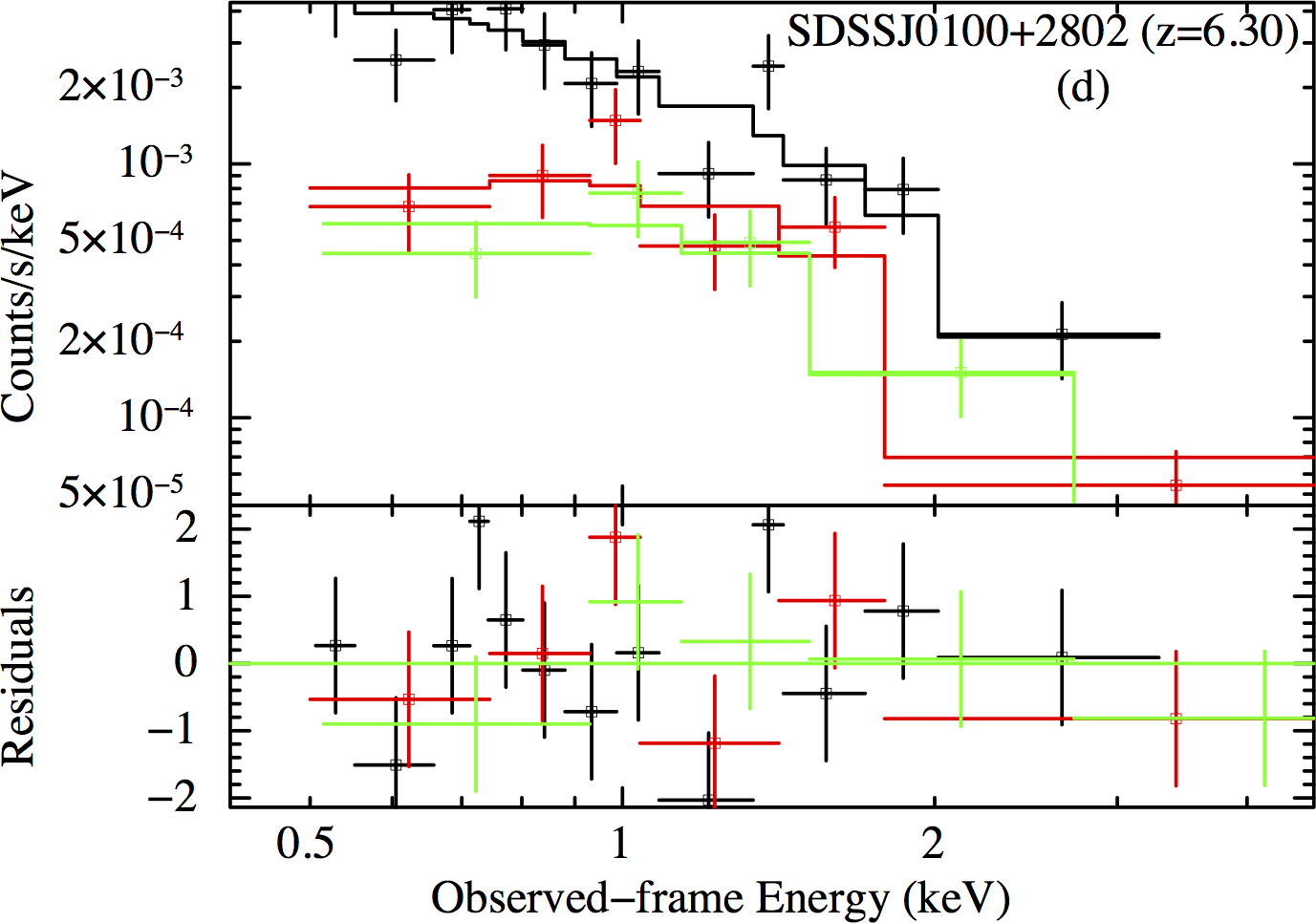}
 \par \vspace{0.5cm}
 \includegraphics[height=6.2cm, width=8.9cm]{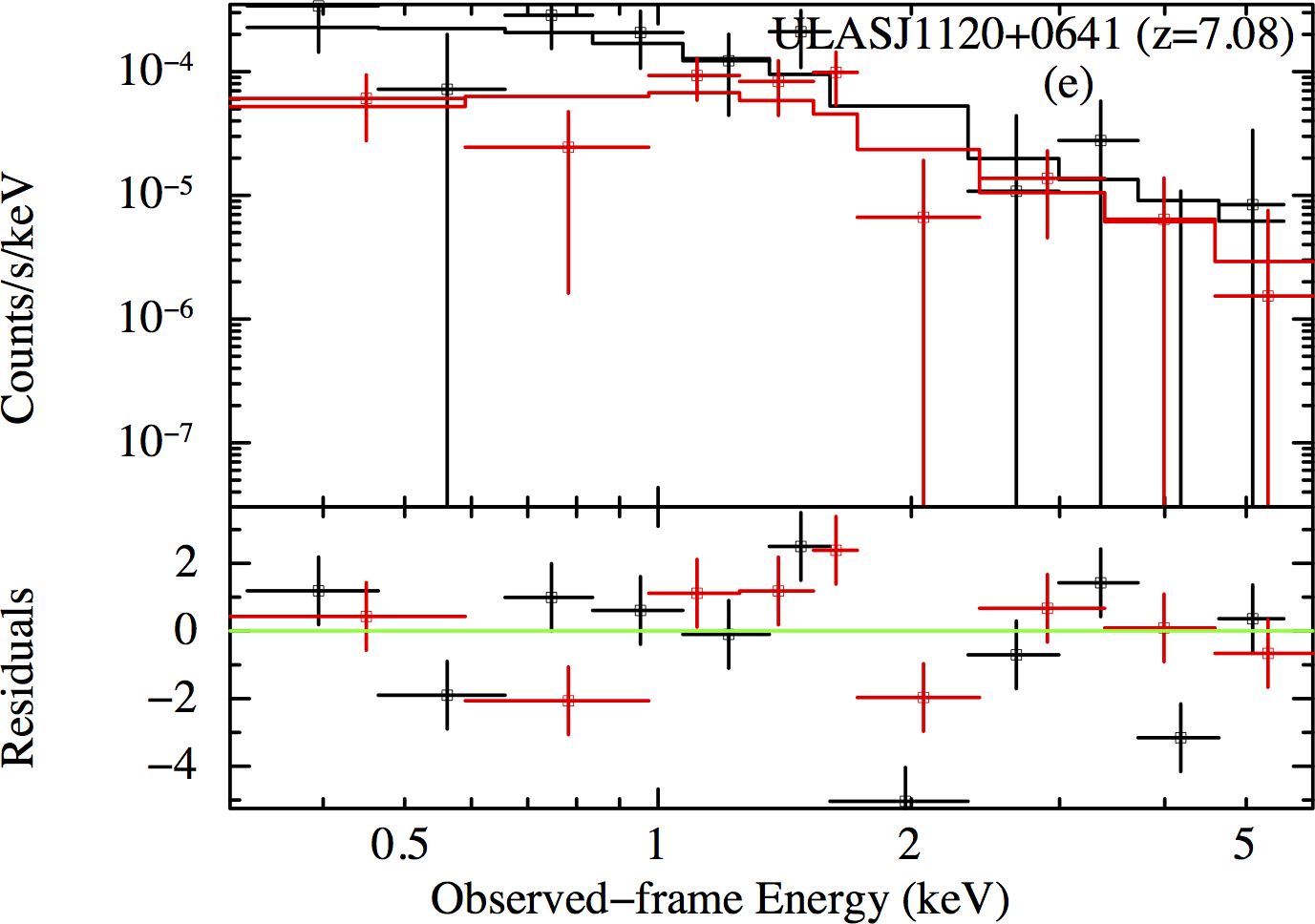}
 \caption{Spectra of the five AGN with the best counting statistics (34 $\le$ net counts $\le$ 148). Spectra in panels \textit{a} and \textit{c} are extracted from \textit{Chandra}, while spectra in panels \textit{b}, \textit{d} and \textit{e} are from XMM\textit{-Newton}. The \textit{a}, \textit{b} and \textit{d} spectra are fit by a power-law model with Galactic and intrinsic absorption; the \textit{c} and \textit{e} spectra are fit by a simple power-law model plus Galactic absorption. The black, red, and green points and lines in the \textit{b} and \textit{d} panels correspond to pn, MOS1, and MOS2, respectively. The black and red points and lines in the \textit{e} panel correspond to pn and combined MOS1 and MOS2, respectively. For the sake of clarity, we display the spectra using a minimum binning of 20, 15, 3, 3 and 20 counts for each bin for J0100+2802, J1030+0524, J1120+0641, J1148+5251 and J1306+0356,  respectively.}
\end{figure*}

\section{Mean X-ray properties of the most distant quasars}

Obtaining accurate values of the X-ray spectral properties, such as the power-law photon index and the intrinsic absorption column density, for most individual sources in this work is hindered by the small numbers of detected photons. To date, only five QSOs at $z$ > 5.5 (the five presented in \S 3.1) have sufficient counting statistics that allow accurate measurements of the \linebreak X-ray spectral properties (Farrah et al. 2004; Moretti et al. 2014; Page et al. 2014; \citealt{Gal17}). Our knowledge of the X-ray spectral properties of quasars at $z$ > 5.5 therefore relies mainly on the joint spectral fitting of samples of these sources (Vignali et al. 2005; Shemmer et al. 2006; Just et al. 2007).
We first selected the 16 quasars detected by \textit{Chandra} and
made a joint spectral fitting analysis using 15 QSOs, excluding the data-set with $\sim$100 net counts of J1306+0356 (but keeping its \linebreak data-set with more limited statistics) and the spectrum of J1148+5251 due to their relatively high statistics (> 30 net counts). 
In all fits we used the Cash statistic and the errors are reported at the 90\% confidence level.
We fit these 15 spectra with a power-law model and associated its value of redshift and Galactic absorption to each source.
We found a mean photon index $\Gamma = 1.93_{-0.29}^{+0.30}$ (C-stat = 223.1 for 151 d.o.f.), which is consistent with those found in previous works.
As a further test, we stacked all the \textit{Chandra} spectra from the detected sources with similar redshift, obtaining two combined spectra, one from sources with 5.7 $\le$ $z$ $\le$ 6.1 (10 QSOs) and one from sources with 6.2 $\le$ $z$ $\le$ 6.5 (5 QSOs), excluding the spectrum of J1120+0641 from the sum, because of its very high redshift, and the data-set of J1306+0356 with a high number of counts. This separation into two redshift bins limits errors caused by summing spectral channels that correspond to different rest-frame energies.

The lower-redshift stack has an average redshift of $z$ = 5.92 and 130 net counts. The one at higher redshift has an average redshift of $z$ = 6.30 and 66 net counts.
We used XSPEC to fit the two spectra with a simple power-law\footnote{In these cases we included a Galactic absorption component, which was kept fixed at a mean N$_H$ value during the fit.} and derived a mean photon index of $\Gamma = 1.92_{-0.27}^{+0.28}$ (C-stat = 48.3 for 91 d.o.f.) for the lower redshift spectrum (Figure 5, left). This value is consistent with the mean photon indices obtained by jointly fitting spectra of luminous and unobscured quasars at lower redshift (1 $\le$ $z$ $\le$ 5.5; e.g., Vignali et al. 2005; Shemmer et al. 2006; Just et al. 2007) and is also consistent with the values predicted by theory (the power-law spectrum is produced by inverse Compton processes caused by interaction of hot-corona electrons with optical/UV photons from the accretion disk; typical values are $\Gamma \sim 1.8 - 2.1$; \citealt{HM91}, 1993). In Figure 5 (right panel) we report the mean photon indices for QSO samples at different redshifts derived from joint fitting or stacking analysis. We did not find any significant evolution of the AGN photon index with redshift up to $z$ $\sim$ 6 and the only two values measured at higher redshift (J1030+0524 at $z$ = 6.31 by Farrah et al. 2004 and J1120+0641 at $z$ = 7.08 by Moretti et al. 2014) are consistent with this non evolutionary trend.

We note that, at $z$ $\sim$ 6, we are sampling rest frame energies in the range 3.5-49 keV. In this band, a hardening of AGN spectra is often observed because of the so called "Compton-reflection hump", that is, radiation from the hot corona that is reprocessed by the accretion disk, which peaks at $\sim$30 keV. However, the mean photon index we derived does not differ from the typical value of Type 1 AGN, suggesting that the presence of the Compton-reflection component is not significant in our sample, as indeed is observed
for luminous QSOs (e.g., \citealt{Pag05}; \citealt{She08}). The individual photon indices we derived in \S 3.1 for the five sources with > 30 net counts
are also consistent with typical values of luminous unobscured QSOs, again suggesting negligible Compton reflection.
For the higher-redshift spectrum we obtained a photon index with poorer constraints than the previous one, $\Gamma = 1.73_{-0.40}^{+0.43}$ (C-stat = 51.0 for 55 d.o.f.), because of the smaller number of counts. This spectrum is characterized by a flatter power-law slope due to the presence of J1148+5251, that has a flatter photon index (see Gallerani et al. 2017). However, this value is still consistent, within the errors, with those present in the literature.
\begin{figure*}
 \centering
 \includegraphics[height=6.2cm, width=8.9cm]{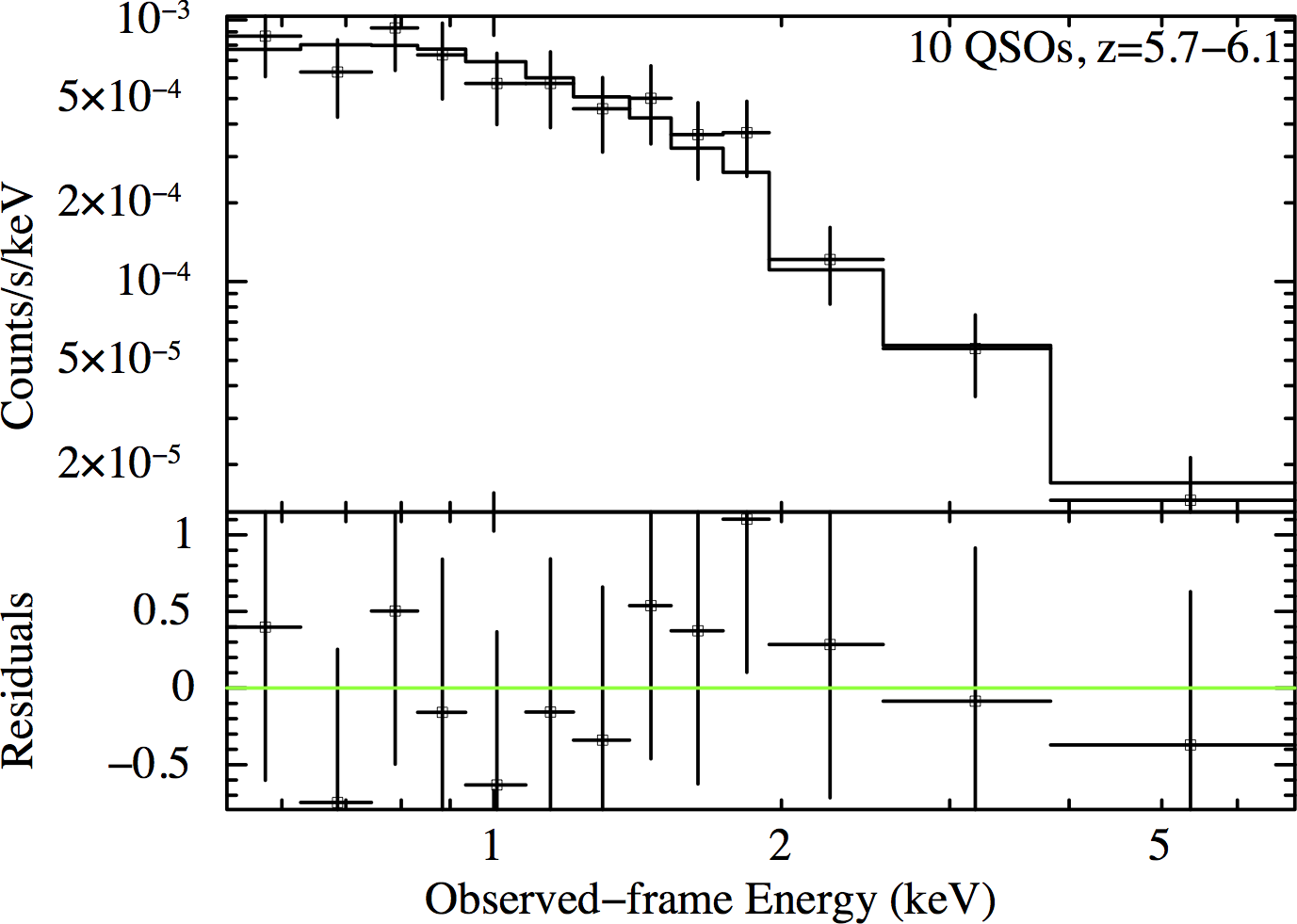} \quad \includegraphics[height=6.5cm, width=9cm]{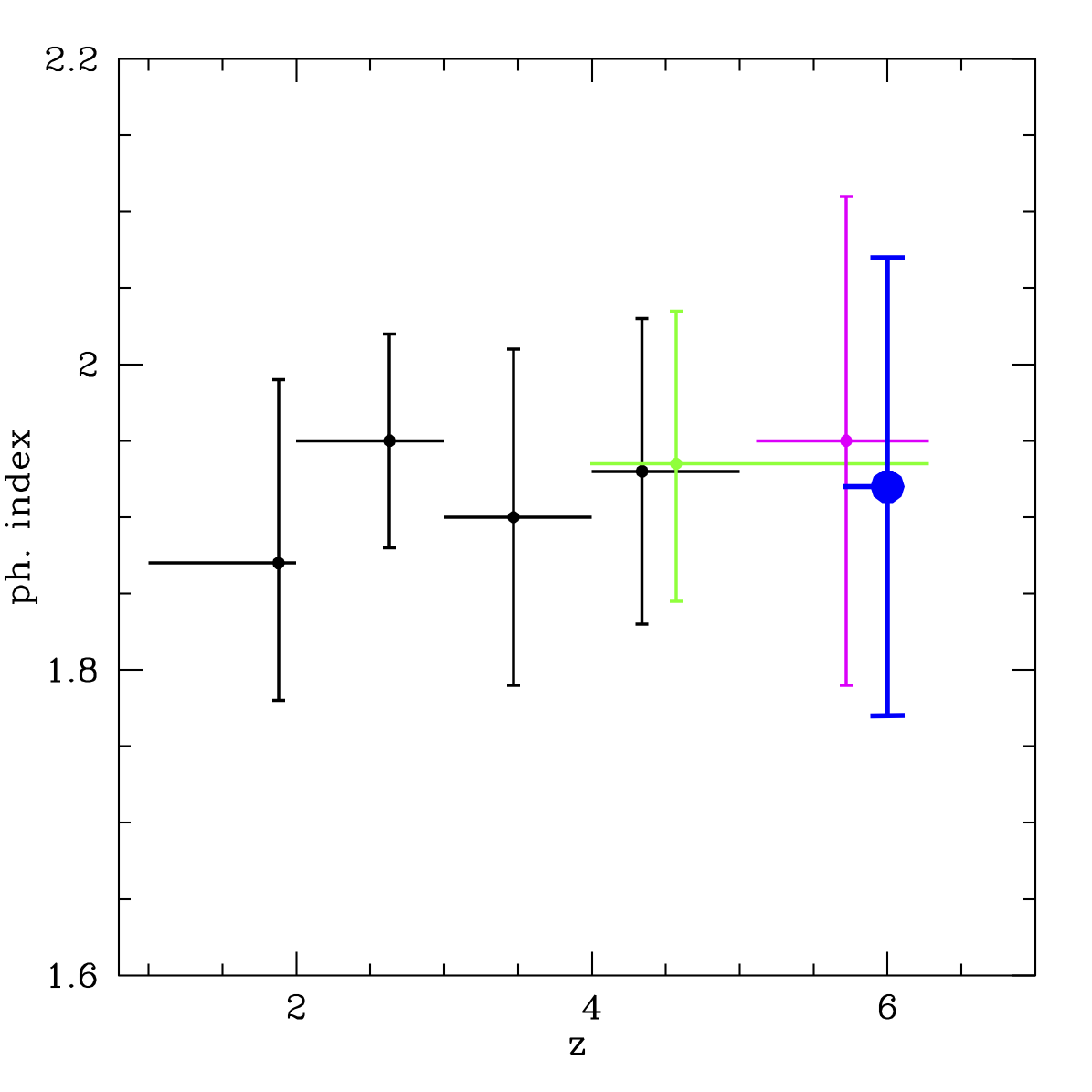}
 \caption{\textit{Left panel}: stacked spectrum of the ten QSOs with 5.7 $\le$ $z$ $\le$ 6.1 fit with a power-law model. On the bottom we report the residuals (data$-$model).
 For clarity we adopted a minimum binning of ten counts per bin.
 \textit{Right panel}: photon index vs. redshift. Black, green, and magenta points are the results of the Just et al (2007), Vignali et al. (2005), and Shemmer et al. (2006) stacking analyses, respectively. The blue point is the photon index of the stacked spectrum shown in the left panel. In all cases the assumed model is a simple power-law and errors are reported at the 68\% confidence level.}
\end{figure*}
Then we fit the two spectra with an absorbed power-law model and $\Gamma$ frozen to 1.9. We found that $N_H \le 8.9\cdot 10^{22}$~cm$^{-2}$ for the former spectrum and $N_H \le 5.0\cdot 10^{23}$~cm$^{-2}$ for the latter spectrum. The limits on the mean column densities are consistent with the values found in the literature and indicate that the population of $z$ > 5.5 luminous QSOs is not significantly obscured, as expected according to their optical and NIR classification. 
Finally, we combined all the spectra used in the two stacking analyses, excluding J1148+5251, and fit them with a power-law model, obtaining a spectrum with 157 net counts and with $\Gamma = 1.83_{-0.24}^{+0.25}$ (C-stat = 62.7 for 108 d.o.f.), fully consistent with the values previously reported.

\section{X-ray and optical properties of the sample}

In Table 3 we provide all the X-ray properties we derived as well as all the optical information available in the literature for our sample.
The details of the Table columns are provided below.

\textit{Column (1)}. - The name of the quasar taken from Ba\~{n}ados (2015) and Ba\~{n}ados et al. (2016).

\textit{Column (2)}. - The monochromatic apparent AB magnitude at the rest-frame wavelength $\lambda$ = 1450 \AA $\:$taken from Ba\~{n}ados (2015).

\textit{Column (3)}. - The absolute magnitude at the rest-frame wavelength $\lambda$ = 1450 \AA $\:$and computed from m$_{1450}$.

\textit{Column (4)}. - The 2500 \AA $\;$rest-frame luminosity, computed from the magnitudes in column (2), assuming a UV-optical power-law slope of $\alpha = -0.5$ (e.g., Shemmer et al. 2006; Just et al. 2007).

\textit{Column (5)}. - The Galactic absorption-corrected flux in the observed-frame 0.5-2.0 keV band. Fluxes were computed using XSPEC for detected sources with > 10 net counts and using PIMMS\footnote{For each \textit{Chandra} observation we set the response to that of the corresponding observing Cycle to account for the effective area degradation.} for QSOs with < 10 net counts and for those undetected (assuming a power-law with $\Gamma = 1.9$). Upper limits are at the 3$\sigma$ level.

\textit{Column (6)}. - The luminosity in the rest-frame 2-10 keV band.

\textit{Column (7)}. - The optical-X-ray power-law slope defined as
\begin{equation}
\alpha_{ox} = \frac{log(f_{2 \: keV}/f_{2500 \: \AA})}{log(\nu_{2 \: keV}/\nu_{2500 \: \AA})},
\end{equation}
where $f_{2 \: keV}$ and $f_{2500 \: \AA}$ are the flux densities at rest-frame 2 keV and 2500 \AA, respectively. The errors on $\alpha_{ox}$ were computed following the numerical method described in \S 1.7.3 of \citealt{Lyo91}, taking into account the uncertainties in the X-ray counts and an uncertainty of 10\% in the 2500 \AA $\;$flux corresponding to a mean $z$-magnitude error of $\sim0.1$.

\textit{Column (8)}. - Upper limits on the column density derived from the spectral fitting for sources with > 10 net counts with a power-law model with $\Gamma$ frozen to 1.9.

In Figure 6 we report the 0.5-2.0 keV flux versus apparent magnitude at 1450 \AA.
\begin{figure}
 \includegraphics[height=10cm, width=8cm, keepaspectratio]{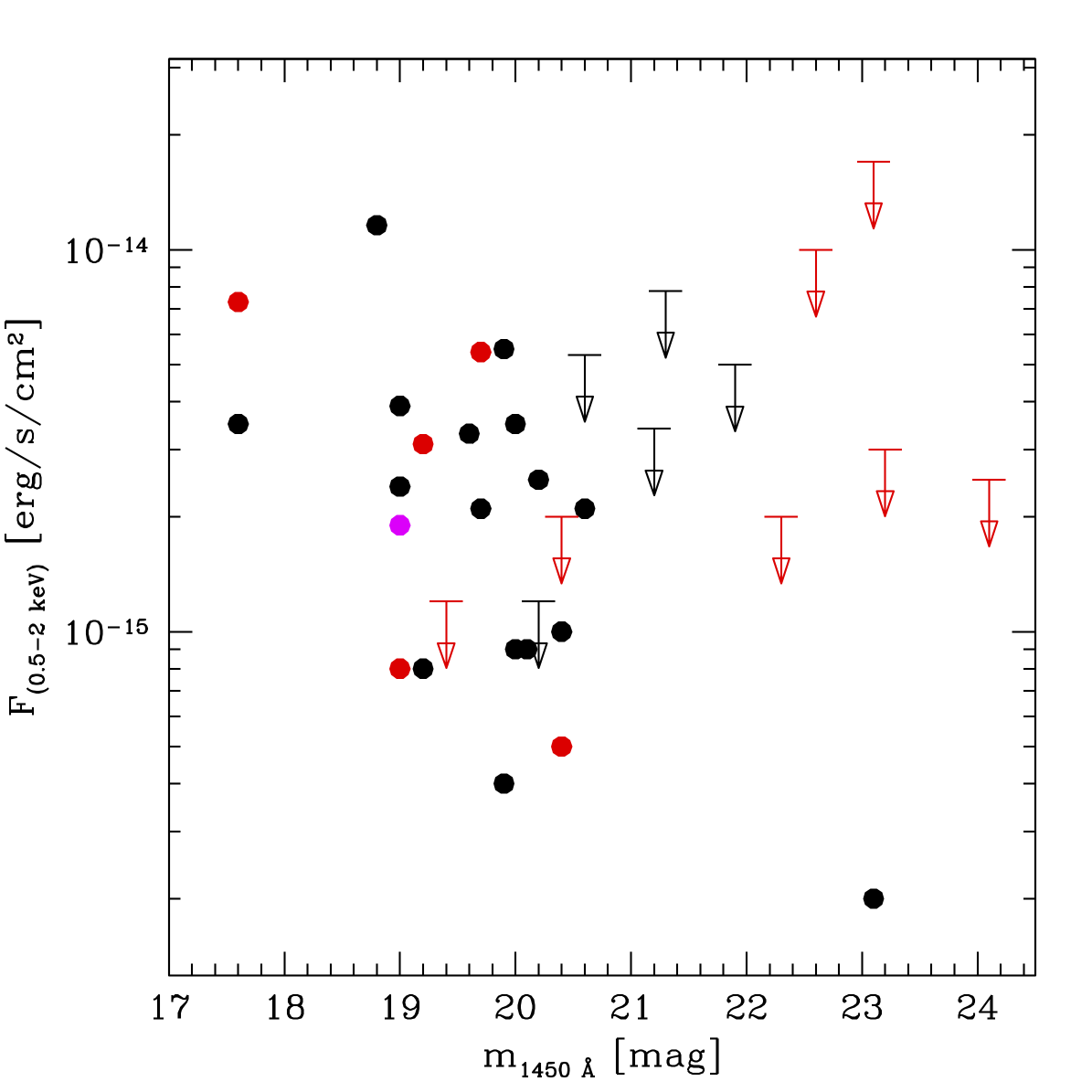}
 \caption{0.5-2 keV flux vs. apparent magnitude at 1450 \AA. Black points indicate the 21 QSOs observed with \textit{Chandra}, red points represent the 12 QSOs observed with XMM and the magenta point is the source observed with \textit{Swift}-XRT. Detected QSOs are shown as filled circles while downward-pointing arrows represent 3$\sigma$ upper limits.}
\end{figure}

\begin{table*}
  \centering
  \captionsetup{justification=centering, labelsep = newline}
      \caption[]{Optical and X-ray properties}
      \begin{adjustbox}{center, max width=\textwidth}
         \begin{tabular}{c c c c c c c c}
            \hline
            \hline \rule[0.7mm]{0mm}{3.5mm}
            \multirow{2}{*}{Object} & \multirow{2}{*}{m$_{1450 \: \AA}$} & \multirow{2}{*}{M$_{1450 \: \AA}$} & log($\nu L_{\nu}$) & $f_x$ & $L$ & \multirow{2}{*}{$\alpha_{ox}$} & \multirow{2}{*}{$N_H$} \\
            & & & (2500 \AA)& 0.5$-$2 keV & 2$-$10 keV & & \\
           (1) & (2) & (3) & (4) & (5) & (6) & (7) & (8) \\
            \hline \rule[0.7mm]{0mm}{3.5mm}
            J1427+3522 & 21.9 & $-$24.7 & 45.5 & < 5.0 & < 2.0 & < $-$1.27 & $-$\\
            \rule[0.7mm]{0mm}{3.5mm}
            J1148+5253 & 23.1 & $-$23.5 & 45.1 & 0.2$_{-0.1}^{+0.2}$ & 0.1$_{-0.05}^{+0.09}$ & $-$1.65$_{-0.12}^{+0.12}$ & $-$\\  
            \rule[0.7mm]{0mm}{3.5mm}
            " $^a$ & " & " & " &  < 17.0 & < 7.1 & < $-$0.89 & $-$\\
            \rule[0.7mm]{0mm}{3.5mm}
 	   J0129-0035 $^a$ & 22.3 & $-$24.3  & 45.4  & < 2.0 & < 0.9 & < $-$1.37 & $-$ \\   
	   \rule[0.7mm]{0mm}{3.5mm}
  	   J1044$-$0125 $^a$ & 19.2 & $-$27.4 & 46.7 & 3.1$_{-0.4}^{+0.5}$  & 1.0$_{-0.1}^{+0.2}$ & $-$1.77$_{-0.02}^{+0.02}$ & < 2.0\\   
	   \rule[0.7mm]{0mm}{3.5mm}
  	   J0836+0054 & 18.8 & $-$27.8 & 46.8 & 11.6$_{-3.8}^{+2.7}$ & 4.2$_{-1.4}^{+1.0}$ & $-$1.61$^{+0.03}_{-0.06}$ & < 1.4 \\ 
	   \rule[0.7mm]{0mm}{3.5mm}
  	  J0002+2550  & 19.0 & $-$27.6 & 46.8 & 3.9$_{-1.6}^{+2.4}$ & 1.7$_{-0.7}^{+1.0}$ & $-$1.76$_{-0.09}^{+0.08}$ & $-$\\ 
	  \rule[0.7mm]{0mm}{3.5mm}
 	  J0840+5624 & 20.0 & $-$26.7 & 46.4 & 0.9$_{-0.5}^{+0.7}$ & 0.4$_{-0.2}^{+0.3}$ & $-$1.85$^{+0.10}_{-0.13}$ & $-$\\
	  \rule[0.7mm]{0mm}{3.5mm}
  	  J0005$-$0006 & 20.2 & $-$26.5 & 46.3 & 2.5$_{-1.9}^{+2.7}$ & 1.5$_{-0.6}^{+1.5}$ & $-$1.65$_{-0.22}^{+0.12}$ & < 1.5\\
	  \rule[0.7mm]{0mm}{3.5mm}
	  J1425+3254 & 20.6 & $-$26.1 & 46.1 & < 5.3 & < 2.4 & < $-$1.46 & $-$\\
	  \rule[0.7mm]{0mm}{3.5mm}
 	  J1335+3533 & 19.9 & $-$26.8 & 46.4 & 0.4$_{-0.2}^{+0.4}$ & 0.2$_{-0.1}^{+0.2}$ & $-$1.97$_{-0.12}^{+0.12}$ & $-$ \\
	  \rule[0.7mm]{0mm}{3.5mm}
  	  J1411+1217  & 20.0 & $-$26.7 & 46.4 & 3.5$_{-2.0}^{+2.3}$ & 1.4$_{-0.6}^{+0.6}$ & $-$1.62$_{-0.14}^{+0.08}$ & < 2.8 \\ 
	  \rule[0.7mm]{0mm}{3.5mm}
  	  J328.7339$-$09.5076 $^a$ & 20.4 & $-$26.3  & 46.2 & < 2.0 & < 0.9 & < $-$1.65 & $-$ \\
	  \rule[0.7mm]{0mm}{3.5mm}
 	  J2053+0047 & 21.2 & $-$25.5 & 45.9 & < 3.4 & < 1.6 & < $-$1.45 & $-$ \\ 
	  \rule[0.7mm]{0mm}{3.5mm}
  	  J0148+0600 $^a$ & 19.4 & $-$27.3 & 46.7 & < 1.2 & < 0.6 & < $-$1.52 & $-$ \\
	  \rule[0.7mm]{0mm}{3.5mm}
  	  J007.0273+04.9571 & 20.2 & $-$26.5  & 46.3 & < 1.2 & < 0.6 & < $-$1.78 & $-$ \\  
	  \rule[0.7mm]{0mm}{3.5mm}
  	  J0216$-$0455 $^a$ & 24.1 & $-$22.6 & 44.8 & < 2.5 & < 1.2 & < $-$1.06 & $-$\\ 
	  \rule[0.7mm]{0mm}{3.5mm}
  	  J1306+0356  & 19.6 & $-$27.1  & 46.6 & 3.3$_{-0.7}^{+0.8}$ & 1.2$_{-0.3}^{+0.3}$ & $-$1.70$_{-0.04}^{+0.04}$ & < 2.5\\
	   \rule[0.7mm]{0mm}{3.5mm}
  	   J1630+4012 & 20.6 & $-$26.1 & 46.2 & 2.1$_{-0.7}^{+0.8}$ & 0.9$_{-0.2}^{+0.3}$ & $-$1.62$_{-0.07}^{+0.05}$ & < 1.4\\
	   \rule[0.7mm]{0mm}{3.5mm}
  	   J0303$-$0019  & 21.3 & -25.4 & 45.9 & < 7.8 & < 3.8 & < $-$1.30 & $-$ \\ 
	   \rule[0.7mm]{0mm}{3.5mm}
  	   J1602+4228 & 19.9 & $-$26.8 & 46.5  & 5.5$_{-1.3}^{+1.4}$  & 3.4$_{-0.5}^{+0.6}$ & $-$1.57$_{-0.05}^{+0.04}$ & < 1.5\\ 
	   \rule[0.7mm]{0mm}{3.5mm}
  	   J2216$-$0016 $^a$ & 23.2 & $-$23.5 & 45.2 & < 3.0 & < 1.5 & < $-$1.17 & $-$ \\ 
	   \rule[0.7mm]{0mm}{3.5mm}
  	   J1048+4637 & 19.2 & $-$27.6 & 46.8 & 0.8$_{-0.4}^{+0.8}$ &  0.4$_{-0.2}^{+0.4}$ & $-$2.02$_{-0.12}^{+0.12}$ & $-$ \\ 
	   \rule[0.7mm]{0mm}{3.5mm}
  	   J1623+3112 & 20.1 & $-$26.7 & 46.4 & 0.9$_{-0.4}^{+0.7}$ & 0.5$_{-0.2}^{+0.4}$ & $-$1.84$_{-0.10}^{+0.10}$ & $-$\\ 
	   \rule[0.7mm]{0mm}{3.5mm}
  	   J0100+2802 & 17.6 & $-$29.2 & 47.5 & 3.5$_{-1.8}^{+2.3}$ & 1.4$_{-0.7}^{+0.9}$ & $-$2.01$_{-0.12}^{+0.08}$ & < 2.6 \\  
	   \rule[0.7mm]{0mm}{3.5mm}
	   " $^a$ & " & " & " & 7.2$_{-0.9}^{+0.4}$ & 3.4$_{-0.5}^{+0.4}$ & $-$1.88$_{-0.02}^{+0.01}$ & < 2.1\\
	   \rule[0.7mm]{0mm}{3.5mm}
	   J025.6821$-$33.4627 $^b$ & 19.0 & -27.8 & 46.9 & 1.9$_{-0.4}^{+0.8}$ & 0.7$_{-0.1}^{+0.3}$ & $-$2.02$_{-0.04}^{+0.06}$ & < 6.7 \\
	   \rule[0.7mm]{0mm}{3.5mm}
  	   J1030+0524 & 19.7 & $-$27.1 & 46.6 & 2.1$_{-0.8}^{+1.3}$ & 1.1$_{-0.4}^{+0.7}$ & $-$1.77$_{-0.08}^{+0.08}$ & $-$\\  
	   \rule[0.7mm]{0mm}{3.5mm}
	   " $^a$ & " & " & " & 5.7$_{-1.1}^{+0.9}$ & 2.6$_{-0.4}^{+0.3}$ & $-$1.60$_{-0.03}^{+0.02}$ & < 0.3\\
	   \rule[0.7mm]{0mm}{3.5mm}
  	   J1148+5251 & 19.0 & $-$27.8 & 46.9 & 2.4$_{-0.5}^{+0.6}$ & 1.3$_{-0.5}^{+1.2}$  & $-$1.86$_{-0.04}^{+0.04}$ & < 5.5\\ 
	   \rule[0.7mm]{0mm}{3.5mm}
	   " $^a$ & " & " & " & 0.8$_{-0.5}^{+0.8}$ & 0.5$_{-0.3}^{+0.5}$ & $-$2.03$_{-0.16}^{+0.11}$ & $-$\\
	   \rule[0.7mm]{0mm}{3.5mm}
  	   J0210$-$0456 $^a$& 22.6 & $-$24.2 & 45.5 & < 10.0 & < 5.5 & < $-$0.91 & $-$  \\ 
	   \rule[0.7mm]{0mm}{3.5mm}
  	   J1120+0641 & 20.4 & $-$26.6 & 46.5 & 1.0$_{-0.5}^{+0.8}$ & 0.7$_{-0.4}^{+0.6}$  & $-$1.81$_{-0.14}^{+0.10}$ & $-$ \\
	   \rule[0.7mm]{0mm}{3.5mm}
	   " $^a$ & " & " & " & 0.7$_{-0.2}^{+0.1}$ & 0.4$_{-0.1}^{+0.1}$ & $-$1.92$_{-0.04}^{+0.03}$ & < 0.8 \\[2pt]
            \hline
         \end{tabular}
        \end{adjustbox}
        \begin{tablenotes}\footnotesize
        \item 	(1) Object name. (2) Monochromatic apparent AB magnitude at the rest-frame wavelength $\lambda$ = 1450 \AA. (3) Absolute magnitude at the rest-frame wavelength $\lambda$ = 1450 \AA \;(see \S 5). (4) Luminosity at the rest-frame wavelength of 2500 \AA. (5) Galactic absorption-corrected flux in the observed \linebreak 0.5-2 keV band in units of 10$^{-15}$ erg cm$^{-2}$ s$^{-1}$. Upper limits and errors are reported at the 3$\sigma$ level and 1$\sigma$ level, respectively. (6) Luminosity in the 2.0-10.0 keV rest-frame band in units of 10$^{45}$ erg s$^{-1}$. (7) The optical-X-ray power-law slope. (8) Upper limits to the column density for detected sources with > 10 net counts in units of 10$^{23}$ cm$^{-2}$.\\
 (a) Sources observed by XMM. X-ray properties were derived averaging the results obtained for the three detectors (pn, MOS1, MOS2).
 (b) Source observed by \textit{Swift}-XRT.
	 \end{tablenotes}
   \end{table*}

\subsection{Source variability}

The five sources with the highest statistics (\S 3.1) have been observed with \textit{Chandra} and XMM in different years, so we checked if these QSOs have varied their X-ray fluxes over the passing of time. J0100+2802, J1030+0524, J1120+0641 and J1148+5251 were observed and detected by both X-ray observatories, and we computed the variability significance using the fluxes reported in Table 3, while RD J1148+5253 is detected only by \textit{Chandra}. For this source the upper limit on the flux derived from XMM data is above the flux value derived from \textit{Chandra} (see Table 3), so there is no clear evidence of variability.
Also J1306+0356 was observed at two different epochs by \textit{Chandra} so, in this case, we computed the variability significance using the fluxes derived from the spectral fit of the two data-sets ($f_{0.5-2 \; keV}=2.7_{-0.3}^{+0.4}$ and $f_{0.5-2 \; keV}=4.5_{-0.5}^{+1.0}$ in units of $10^{-15}$ erg cm$^{-2}$ s$^{-1}$). All the computed significances are below the 2$\sigma$ level, so there is no clear evidence of flux variability in these five sources. These results are consistent with those found for lower redshift sources (4.10 $\le$ $z$ $\le$ 4.35), with comparable X-ray luminosities, and strengthen the idea that the X-ray variability does not increase with redshift (Shemmer et al. in prep.).

\subsection{Multi-band information from the literature}

QSOs with peculiar multi-band emission properties could be characterized by different emission or accretion processes that can also affect their X-ray spectra. For example, 
radio-loud AGN usually have X-ray spectra flatter than radio-quiet QSOs, because of the contribution from the jet (e.g., \citealt{Wu13}).
 Thus, we checked if there are any peculiar QSOs in our sample that also have peculiar X-ray properties linked to their different nature. First, we checked the VLA FIRST catalog (\citealt{Whi97}) and the literature to derive information about the radio loudness (RL) of our sources, adopting the definition by \citet{Kel89}: $RL=f_{\nu,5\:GHz}/f_{\nu,4400\: \AA}$, where $f_{\nu,5\:GHz}$ is the 5 GHz radio rest-frame flux density and $f_{\nu,4400\: \AA}$ is the 4400 $\AA$ optical rest-frame flux density, and a quasar is considered radio loud if RL > 10. Assuming an average optical spectral index of $\alpha = -0.5$, we extrapolated the optical rest-frame flux density at 4400 \AA \;from the \textit{WISE} W$_1$ ($\lambda \sim$ 3.4 $\mu m$) magnitude, when available, or from $m_{1450}$ otherwise. Twenty-five sources have upper limits on their radio fluxes, two have not been observed by VLA (J328.7339-09.5076 and J025.6821-33.4627) and two (J083643.8+005453.2 and J010013.0+280225.9) are detected with a RL $\sim$12 and $\sim$0.3, respectively. The first value is consistent with the one derived by \citet{Bana15} and indicates a moderate level of radio emission that is not supposed to significantly affect its X-ray spectrum (but see \citealt{Mil11}). The two VLA-unobserved sources are also not observed by NVSS. Summarizing, from the values of the derived RL parameters, we found that there are no clear indications of the presence of extreme Radio Loud QSOs in our sample. We also checked in the literature for the presence of any Broad Absorption Line (BALQ), Weak-Line (WLQ) or Weak Infrared QSOs (sources with a weak emission at $\sim$10 $\mu m$ rest-frame due to a possible lack of torus emission component; \citealt{Jia10}).
In our sample five QSOs are classified as BALQs, six as WLQs and two as Weak-IR QSOs (see Table 1). WLQs are defined as quasars having rest-frame equivalent widths (EWs) of < 15.4 $\AA$ for the Ly$\alpha$+N V emission-line complex (\citealt{DS09}). This could be due to either an extremely high accretion rate, that may result in a relatively narrow UV-peaked SED (\citealt{Luo15}) in which prominent high-ionization emission lines are suppressed (the so called Baldwin effect; \citealt{Bal77}), or a significant deficit of line-emitting gas in the broad-emission line region (\citealt{She10}). In our case the WLQ X-ray properties are consistent with those of non WLQs (see Table 3).

\subsection{Comparison of the optical properties with lower redshift results}

The optical-X-ray power-law slope ($\alpha_{ox}$), defined in Equation~1 in \S 5, is expected to trace the relative importance of the disk versus corona.
Previous works have shown that there is a significant correlation between $\alpha_{ox}$ and the monochromatic $L_{2500 \: \AA}$ ($\alpha_{ox}$ decreases as $L_{2500 \: \AA}$ increases; \citealt{Ste06}; \citealt{LR17}), whereas the apparent dependence of $\alpha_{ox}$ on redshift can be explained by a selection bias (\citealt{Zam81}; Vignali et al. 2003; \citealt{Ste06}; Shemmer et al. 2006; Just et al. 2007; \citealt{Lus10}; but see also \citealt{Kel07}).
We further examine the $\alpha_{ox}-L_{2500 \: \AA}$ relationship adding our sample of 29 high-redshift QSOs to previous measurements of $\alpha_{ox}$.
We have plotted $\alpha_{ox}$ versus $L_{2500 \: \AA}$ for all the X-ray quasars of our sample in Figure 7, including 1515 QSOs from lower redshift analyses (X-ray selected: 529 from Lusso et al. 2010, 174 from \citealt{Marchese12}; optically selected: 743 from \citealt{LR16}, 11 from \citealt{Vig03}, 13 from Vignali et al. 2005, 13 from Shemmer et al. 2006, 32 from \citealt{Jus07}). We excluded eight sources from the original sample of Shemmer et al. (2006) because they are also present in our sample (our results for these eight sources are consistent with those derived by Shemmer et al. 2006), obtaining a final sample of 1544 QSOs.
Our sample follows the correlation between $\alpha_{ox}$ and UV luminosity with no detectable dependence on redshift. We performed linear regression on the data (13 of them have upper limits on $\alpha_{ox}$) using the ASURV software package (\citealt{Lav92}), confirming and strengthening the finding in previous studies that $\alpha_{ox}$ decreases with increasing rest-frame UV luminosity. We found the best-fit relation between $\alpha_{ox}$ and $L_{2500 \: \AA}$ to be:
\begin{equation}
  \alpha_{ox} = (-0.155 \pm 0.003)log(L_{2500 \: \AA})+(3.206 \pm 0.103).
\end{equation}
Errors are reported at the 1$\sigma$ confidence level.
This correlation is based on the highest number of QSOs available. These best-fit parameters are consistent with those derived by Just et al. (2007) and by Lusso et al. (2010).
We note that the presence of our and the Shemmer et al. (2006) samples improves coverage at $z$ $\approx$ 5-6, showing that our analysis supports the idea that luminous AGN SEDs have not significantly evolved out to very high redshift. We also obtained a best-fit relation excluding the X-ray selected data (\citealt{Lus10} and \citealt{Marchese12}) and found that is consistent with Equation 2.
\begin{figure}
 \includegraphics[height=10cm, width=8cm, keepaspectratio]{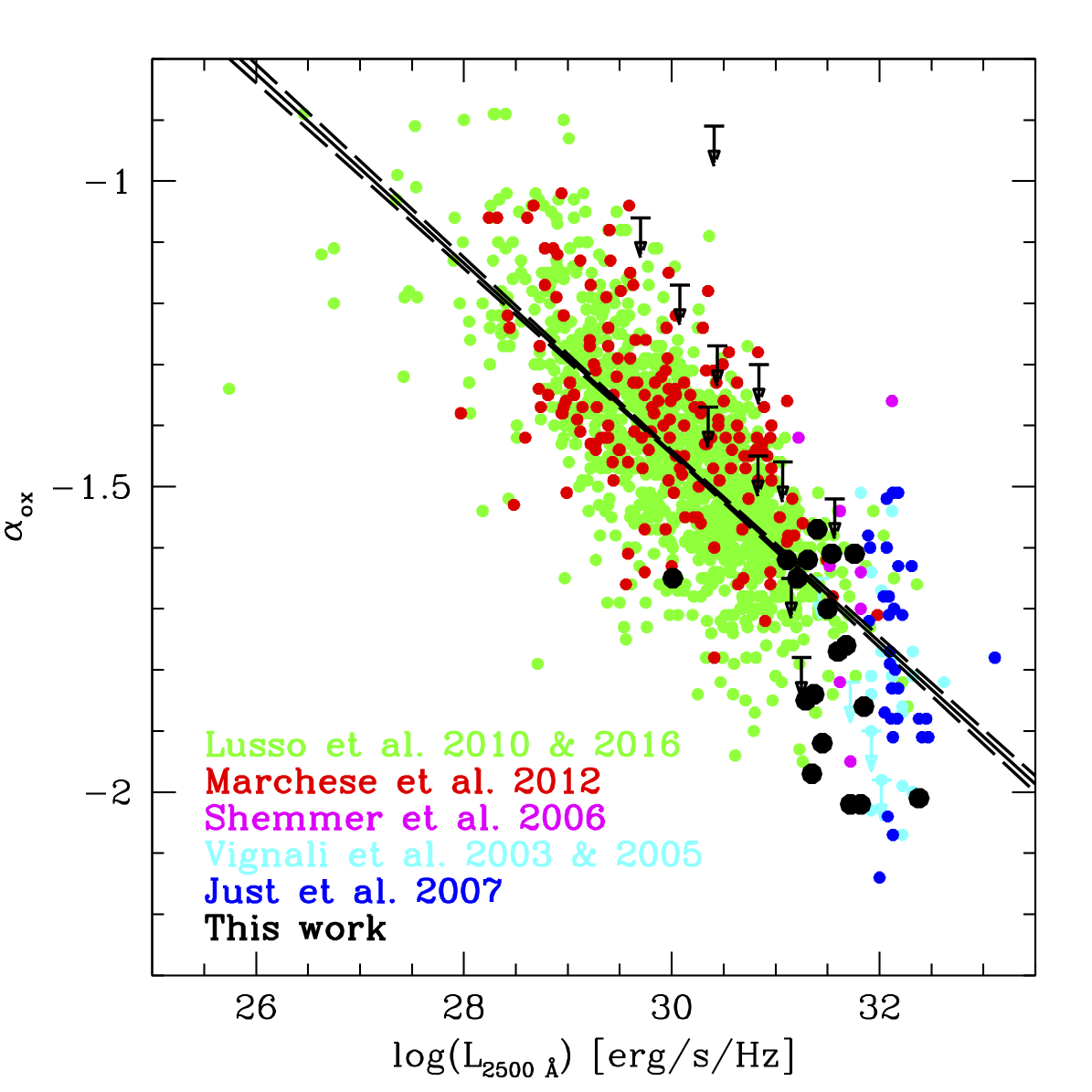}
 \caption{$\alpha_{ox}$ versus UV monochromatic luminosity for 1544 QSOs. Black dots correspond to our detected QSO sources, downward-pointing arrows represent undetected sources. Colored points correspond to different literature samples as labeled. The black solid line is the best-fit relation found in our work, while the dashed black lines represent the uncertainty in the relation.}
\end{figure}

\section{Summary and conclusions}

We made a complete and uniform study of the X-ray properties of the most-distant quasars at $z$ > 5.5. This is the most up-to-date analysis of the X-ray properties of early AGN. Our main results are the following:

\begin{itemize}
	\item We started from a parent sample of 198 spectroscopically confirmed QSOs at $z$ > 5.5 and considered the 29 objects that have been observed by \textit{Chandra}, XMM\textit{-Newton}, and \textit{Swift}-XRT. Eighteen of them are detected in the X-ray band (0.5-7.0 keV). \\
	\item Five sources have sufficient counting statistics (> 30 net counts) to allow us to fit their spectra with a power-law model with $\Gamma$ free to vary. For these quasars we obtained values of the photon index $\Gamma \sim 1.6-2.4$ consistent with those present in literature (Farrah et al. 2004; Moretti et al. 2014; Gallerani et al. 2016) and those expected from theory (Haardt \& Maraschi 1993). \\
	\item By performing a spectral stacking analysis we derived the mean photon index of the early AGN population. We divided our 15 \textit{Chandra} detected sources into two redshift bins: 5.7 $\le$ $z$ $\le$ 6.1 (10 sources) and 6.2 $\le$ $z$ $\le$ 6.5 (5 sources). We obtain $\Gamma = 1.92_{-0.27}^{+0.28}$ for the first stacked sub-sample and $\Gamma = 1.73_{-0.40}^{+0.43}$ for the second one. We do not find a significant change in $\Gamma$ with cosmic time over the redshift range $z \approx 1.0-6.4$. This means that, similarly to optical properties (e.g., Mortlock et al. 2011; Barnett et al. 2013), also the X-ray spectral properties of luminous QSOs do not significantly evolve over cosmic time.
The upper limits to the mean column density derived from the stacking analysis are $N_H < 8.9 \cdot 10^{22}$ cm$^{-2}$ for the first sub-sample and $N_H < 5.0 \cdot 10^{23}$ cm$^{-2}$ for the second one, showing that these luminous high-redshift QSOs are not significantly obscured, as expected from their optical classification as Type 1 AGN.\\
	\item Combining our sample with literature works, we confirmed that, by using a statistically larger sample, the $\alpha_{ox}$ parameter depends on UV monochromatic luminosity. The X-ray-to-optical flux ratios of luminous AGN have not significantly evolved up to $z$ $\sim$ 6.
\end{itemize}

\begin{acknowledgements}
 We acknowledge financial contribution from the agreement ASI-INAF I/037/12/0. W. N. B. thanks Chandra X-ray Center grant GO5-16089X and the
Willaman Endowment for support. We thank G. Risaliti and E. Lusso for useful discussions.
\end{acknowledgements}

\bibliographystyle{aa}
 \bibliography{riccardo}

\end{document}